\documentclass{elsart}

\usepackage{graphicx}

\usepackage{amssymb}

\newcommand{\beq}{\begin{equation}}
\newcommand{\eeq}{\end{equation}}
\newcommand{\E}{\mathcal{E}}

\begin{document}

\begin{frontmatter}



\title{An Unsplit Godunov Method for Ideal MHD via Constrained Transport}


\author[label1]{Thomas A. Gardiner}
\author[label1,label2]{\& James M. Stone}

\address[label1]{Department of Astrophysical Sciences \\ 
Princeton University \\
Princeton, NJ 08544}
\address[label2]{Program in Applied and Computational Mathematics \\
Princeton University \\
Princeton, NJ 08544}

\begin{abstract}
We describe a single step, second-order accurate Godunov scheme for
ideal MHD based on combining the piecewise parabolic method (PPM) for
performing spatial reconstruction, the corner transport upwind (CTU)
method of Colella for multidimensional integration, and the constrained
transport (CT) algorithm for preserving the divergence-free constraint
on the magnetic field.  We adopt the most compact form of CT, which
requires the field be represented by area-averages at cell faces.
We demonstrate that the fluxes of the area-averaged field used by CT
can be made consistent with the fluxes of the volume-averaged field
returned by a Riemann solver if they obey certain simple relationships.
We use these relationships to derive new algorithms for constructing the
CT fluxes at grid cell corners which reduce
exactly to the equivalent one-dimensional solver for plane-parallel,
grid-aligned flow.  We show that the PPM reconstruction algorithm must
include multidimensional terms for MHD, and we describe
a number of important extensions that must be made to CTU in order for
it to be used for MHD with CT.  We present the results of a variety of
test problems to demonstrate the method is accurate and robust.

\end{abstract}

\begin{keyword}

\PACS 
\end{keyword}
\end{frontmatter}

\section{Introduction}
\label{sec:intro}

\par
In recent years a variety of numerical algorithms for multidimensional
magnetohydrodynamics (MHD) based on Godunov's method have been
developed
\cite{Balsara-Spicer,Crockett,Dai-Woodward-97,Falle98,Londrillo-DelZanna04,Pen-MHD-03,RMJF98}.
There are two important extensions to the basic hydrodynamical
algorithm that are required for MHD.  The first is an extension of the
Riemann solver used to compute the fluxes of each conserved quantity
to MHD; the second is a method by which the divergence free constraint
${\bf\nabla\cdot B} = 0$ is imposed upon the numerically evolved
magnetic field.  Of the two, the latter has emerged as the more
difficult to achieve.

\par
The fact that it is important to ensure the numerically evolved field
satisfies the divergence free constraint was first noted by Brackbill
\& Barnes \cite{Brackbill-Barnes}. They pointed out that the Lorentz
force is not orthogonal to ${\bf B}$ if ${\bf \nabla \cdot B} \neq 0$,
and that this could lead to incorrect dynamics in a well defined test
problem.  More recently T\'oth \cite{Toth-divB} has shown that in some
circumstances it is possible to get the wrong jump conditions across
MHD shocks if the constraint is not satisfied (this is also evident
in the method of \cite{Falle98}).

\par 
Currently, there are three methods by which the divergence free
constraint is applied in Godunov schemes.  The first is to use a Hodge
projection to clean the magnetic field of any divergence after each
time step (e.g. \cite{Balsara-Kim,Crockett,Toth-divB}).  The second is
to extend the system of conservation laws with an evolutionary
equation for the divergence designed to minimize the accumulation of
error in any one location.  Examples of this approach include the
eight-wave scheme \cite{Powell99} and the GLM-MHD scheme
\cite{Dedner02}.  Finally, the third is to design the difference
equations for the magnetic field to explicitly conserve magnetic flux,
and so preserve the divergence free constraint.  The latter method
termed constrained transport (CT) by \cite{Evans-Hawley} has proved
successful in other MHD algorithms \cite{Clarke96,Stone-Norman-92b}
and is the method adopted here.

\par
The most compact CT difference formulae are built upon area-averaged
magnetic field components located at the faces of a grid cell, rather than
volume-averaged field components located at grid cell centers (CT
algorithms built upon cell-centered fields have been developed in
\cite{Toth-divB}, however they require averaging over a stencil
which is larger than that used to compute the fluxes).  The need for a
staggered grid is often thought of as a disadvantage of CT.  In fact,
however, it reflects one of the most attractive properties of CT: the
fundamental conserved property of the magnetic field in MHD is the
magnetic flux (which is an area- rather than volume-average) and by
design, CT conserves the magnetic flux (and therefore it preserves the
divergence free character of the magnetic field) in an integral sense,
over the smallest discretization scale, the grid cell size.

\par
We would like to combine CT with a finite volume shock capturing method.
It may at first seem inconsistent to build a numerical algorithm
which mixes a finite volume approach (which conserves integrals
of volume-averaged values) with CT (which conserves integrals of
area-averaged magnetic fluxes at grid cell faces).  In this paper we
show that, provided the fluxes of the volume- and area-averaged fields
obey certain simple relationships, the finite-volume and CT approaches
can be made consistent (see also \cite{Londrillo-DelZanna04}).  Most
importantly, the relationships we derive allow us to develop algorithms
for constructing fluxes of the face-centered field (located at grid cell
corners in 2D) from the fluxes of the volume-averaged field computed
by a Riemann solver (located at grid cell faces).  We demonstrate that
multidimensional algorithms developed in this way reduce exactly to the
equivalent one-dimensional solver for plane-parallel, grid-aligned flow.
This is one difference between the methods developed here and previous
implementations of CT in Godunov schemes 
{}\cite{Balsara-Spicer,Dai-Woodward-98,RMJF98}.

\par
We combine the CT algorithms developed in this paper with the
piecewise parabolic method (PPM) \cite{Colella-Woodward-PPM} using
Roe's linearization as the MHD Riemann solver \cite{Cargo-Gallice}.
An essential ingredient of PPM is a spatial reconstruction step to
compute time-advanced estimates of the conserved variables at grid
faces.  In this paper we show that for MHD, this reconstruction step
must include multidimensional terms in the induction equation (used to
reconstruct the transverse components of the field).  We argue that
dimensionally split MHD algorithms cannot preserve the divergence free
constraint between each one-dimensional update, therefore we adopt the
unsplit corner transport upwind (CTU) algorithm of Colella
\cite{Colella-CTU} to develop a multidimensional algorithm.  However,
there are a number of important extensions that must be made to CTU to
make it suitable for MHD with CT.  These include using a CT update for
the magnetic field during the predict step, and inclusion of
multidimensional terms along with the transverse flux gradients used
to predict multidimensional fluxes.  We describe these extensions in
detail.

\par
The resulting two-dimensional MHD PPM algorithm uses a single step
update, is second order accurate, and is fully conservative.  Hence it
is ideally suited for use on a statically or adaptively refined mesh.
Moreover, the two-dimensional algorithm reduces exactly to the base
one-dimensional algorithm for planar, grid-aligned flows.  Although
this paper will only describe the combination of CT with CTU, there is
no reason why the CT algorithm described here could not be combined
with other unsplit methods.  To simplify the discussion we confine
ourselves to a two-dimensional algorithm in this paper.

\par
The paper is organized as follows.  In \S\ref{sec:CT-FV} we review the
finite volume and CT methods in order to demonstrate the relationships
between the area- and volume-averaged magnetic fields and their
fluxes.  In \S\ref{sec:ss_ct_Godunov} we use these relationships to
derive algorithms for constructing the fluxes of face-centered
area-averaged fields needed by CT from the fluxes of cell-centered
volume-averaged fields returned by a Riemann solver.  We present a
single step first-order Godunov method and use it to test different
algorithms for constructing the fluxes used in CT.  This method forms
the first half of the second-order CTU + CT integration algorithm
developed in section \S\ref{sec:ctu_ct_Godunov}.  In \S\ref{sec:tests}
we present a variety of tests which demonstrate the linear and
nonlinear behavior of the scheme.  Finally, in \S\ref{sec:Conclusions}
we conclude.

\section{Constrained Transport in Finite Volume Schemes}
\label{sec:CT-FV}

The equations of ideal magnetohydrodynamics (MHD) can be written in
conservative form as
\begin{eqnarray}
\frac{\partial \rho}{\partial t} + 
{\bf\nabla\cdot} \left(\rho{\bf v}\right) & = & 0 
\label{eq:cons_mass} \\
\frac{\partial \rho {\bf v}}{\partial t} + 
{\bf\nabla\cdot} \left(\rho{\bf vv} - {\bf BB}\right) +
{\bf \nabla} P^* & = & 0 \\
\frac{\partial {\bf B}}{\partial t} + 
{\bf\nabla\cdot} \left({\bf v B - B v}\right) & = & 0 \\
\frac{\partial E}{\partial t} + 
\nabla\cdot((E + P^*) {\bf v} - {\bf B} ({\bf B \cdot v})) & = & 0 
\label{eq:cons_energy}
\end{eqnarray}
where $\rho$ is the mass density, $\rho{\bf v}$ is the momentum
density, ${\bf B}$ is the magnetic field, and $E$ is the total energy
density.  The total pressure $P^* \equiv P + ({\bf B \cdot B})/2$ where $P$
is the gas pressure, and the total energy density $E$ is related to the
internal energy density $\epsilon$ via
\beq
E \equiv \epsilon + \rho({\bf v \cdot v})/2 + ({\bf B \cdot B})/2 ~.
\eeq
Throughout this paper we will assume an ideal gas equation of state
for which $P = (\gamma - 1) \epsilon$, where $\gamma$ is the ratio of
specific heats.  Unless otherwise stated, we take $\gamma=5/3$.  None
of the main results described in this paper depend directly upon the
equation of state.  Note also that we have chosen a system of units in
which the magnetic permeability $\mu=1$.

\par
In addition to the evolutionary conservation laws equations
(\ref{eq:cons_mass}) through (\ref{eq:cons_energy}), the magnetic
field must also obey the divergence free constraint,
i.e. ${\bf\nabla\cdot B}=0$.  It is of paramount importance that the
numerically evolved field satisfy this constraint at all times,
otherwise, for example, the system of equations for the conservative
variables is inconsistent with the same system written in terms of the
primitive variables, i.e. $(\rho,~{\bf v},~{\bf B},~P)$.

\par
The algorithm described in this paper is built upon finite volume (FV)
methods, in which the conserved variables are averaged over grid cell
volumes.  On the other hand, the CT method is built upon
area-averaging of the magnetic field, leading to difference equations
for the magnetic flux through the surfaces of grid cells.  In the
following subsections, we briefly review the FV and CT schemes in
order to arrive at consistent relationships between the volume- and
area-averaged magnetic field and their associated numerical fluxes.

\subsection{Finite Volume Method}

Consider a regular, two dimensional cartesian grid with grid cell
$(i,j)$ centered at $(x_i,y_j)$ and of size $(\delta x,\delta y)$.
The conservative system of equations for ideal MHD can be written in
vector form as
\beq
\frac{\partial q}{\partial t} + {\bf \nabla \cdot f} = 0 
\eeq
where
\beq
q \equiv \left (
\begin{array}{c}
\rho \\
\rho v_x \\
\rho v_y \\
\rho v_z \\
B_x \\
B_y \\
B_z \\
E
\end{array}
\right )
\eeq
is the vector of conserved variables and 
\beq
f_x \equiv \left (
\begin{array}{c}
\rho v_x \\
\rho v_x^2 + P^* - B_x^2 \\
\rho v_x v_y - B_x B_y \\
\rho v_x v_z - B_x B_z \\
0 \\
v_x B_y - B_x v_y \\
v_x B_z - B_x v_z \\
(E + P^*)v_x - B_x({\bf B \cdot v})
\end{array}
\right ) ~,
f_y \equiv \left (
\begin{array}{c}
\rho v_y \\
\rho v_y v_x - B_y B_x \\
\rho v_y^2  + P^* - B_y^2 \\
\rho v_y v_z - B_y B_z \\
v_y B_x - B_y v_x \\
0 \\
v_y B_z - B_y v_z \\
(E + P^*)v_y - B_y({\bf B \cdot v})
\end{array}
\right ) ~,
\eeq
are the flux vectors.
Integrating over the volume of grid cell $(i,j)$ and the time interval
$\delta t = t^{n+1} - t^{n}$ and applying Gauss's theorem we obtain
\beq
q_{i,j}^{n+1} = q_{i,j}^n 
+ \frac{\delta t}{\delta x}
 \left( F_{x,i-1/2,j}^{n+1/2} - F_{x,i+1/2,j}^{n+1/2} \right)
+ \frac{\delta t}{\delta y}
 \left( F_{y,i,j-1/2}^{n+1/2} - F_{y,i,j+1/2}^{n+1/2} \right) 
\label{eq:fv_int}
\eeq
the integral form of the evolution equation.  The conserved quantities
\beq
q_{i,j}^n \equiv \frac{1}{\delta x \delta y} 
\int_{y_i - \delta y/2}^{y_i + \delta y/2}
\int_{x_i - \delta x/2}^{x_i + \delta x/2} 
 q(x,y,t^n) dx dy ~,
\eeq
are averaged over the grid cell volume and the fluxes
\beq
F_{x,i\pm1/2,j}^{n+1/2} \equiv
 \frac{1}{\delta y \delta t} \int_{t^n}^{t^{n+1}}
\int_{y_i - \delta y/2}^{y_i + \delta y/2} f_x (x_i \pm \delta x/2,y,t) dy dt
\label{eq:fv_xflux}
\eeq
\beq
F_{y,i,j \pm 1/2}^{n+1/2} \equiv
 \frac{1}{\delta x \delta t} \int_{t^n}^{t^{n+1}}
\int_{x_i - \delta x/2}^{x_i + \delta x/2} f_y (x,y_i \pm \delta y/2,t) dx dt
\label{eq:fv_yflux}
\eeq
are averaged over the surface area of a grid cell face and the time
interval $\delta t$.  Typically one approximates the flux integrals in
equations (\ref{eq:fv_xflux}) and (\ref{eq:fv_yflux}) to some order of
accuracy, while maintaining strict conservation by evolving the
conserved quantities through equation (\ref{eq:fv_int}).  Note when
written in this form, the components of the flux vectors are non-zero
for the transverse components of the magnetic field only, meaning that
directionally split updates of the volume averaged field based on
these fluxes will not generally satisfy the divergence free constraint
between directional sweeps.  This suggests directionally split algorithms
are inappropriate for MHD.

\subsection{Constrained Transport Method}

\par
In the CT method, the integral form of the induction equation is based
on area rather then volume averages.  Starting from the differential
form of the induction equation,
\beq
\frac{\partial {\bf B}}{\partial t} + {\bf \nabla \times \E} = 0
\label{eq:induct}
\eeq
where the electric field ${\bf \E = -v \times B}$ in ideal MHD, one
may integrate over the bounding surface of a grid cell and use Stoke's
Theorem to obtain
\beq
B_{x,i\pm1/2,j}^{n+1} = B_{x,i\pm1/2,j}^n 
+ \frac{\delta t}{\delta y}
 \left( \E_{z,i\pm1/2,j-1/2}^{n+1/2} - \E_{z,i\pm1/2,j+1/2}^{n+1/2} \right) 
\label{eq:ct_Bx_int}
\eeq
\beq
B_{y,i,j\pm1/2}^{n+1} = B_{y,i,j\pm1/2}^n 
- \frac{\delta t}{\delta x}
 \left( \E_{z,i-1/2,j\pm1/2}^{n+1/2} - \E_{z,i+1/2,j\pm1/2}^{n+1/2} \right) 
\label{eq:ct_By_int}
\eeq
as the integral form of the evolution equation.  The magnetic field
components
\beq
B_{x,i\pm1/2,j}^n \equiv \frac{1}{\delta y} 
\int_{y_i - \delta y/2}^{y_i + \delta y/2}
 B_x(x_i \pm \delta x/2,y,t^n) dy 
\eeq
\beq
B_{y,i,j\pm1/2}^n \equiv \frac{1}{\delta x} 
\int_{x_i - \delta x/2}^{x_i + \delta x/2}
 B_y(x, y_i \pm \delta y/2,t^n) dx 
\eeq
are averaged over the grid cell bounding faces and 
\beq
\E_{z,i\pm1/2,j\pm1/2}^{n+1/2} \equiv
 \frac{1}{\delta t} \int_{t^n}^{t^{n+1}}
\E_z (x_i \pm \delta x/2,y_i \pm \delta y/2,t) dt
\label{eq:ct_Ez_int}
\eeq
is averaged over the time interval $\delta t$.  Note the fundamental
representation of the magnetic field is an area-average at cell faces.
Although CT-like difference formulae are possible based on volume
averaged fields at cell centers \cite{Toth-divB}, they preserve a
discretization of the divergence on a different (larger) stencil than
used to compute the fluxes.

\par
Just as in finite volume methods, one typically approximates the
electric field (flux) integral in equation (\ref{eq:ct_Ez_int}) to
some order of accuracy and applies equations (\ref{eq:ct_Bx_int}) and
(\ref{eq:ct_By_int}) to evolve the magnetic field components in time.
Nevertheless, in a manner exactly analogous to the finite volume
method, conservation of magnetic flux is strictly enforced, implying
that the net magnetic charge interior to a grid cell vanishes at time
$t^{n+1}$ if it did so at time $t^n$.  As such, the preservation of
${\bf \nabla \cdot B} = 0$ (in an integral sense) in the CT method is
as fundamental as, e.g., the conservation of mass in a finite volume
method.  Moreover, in the CT approach, the magnetic field components
are averaged over the smallest dimensional volume necessary so as to
transform the differential equation into its integral form.  In this
way one maintains the maximal ``point-wise'' information possible,
thereby minimizing the dissipation inherent in the averaging process.

\subsection{Consistency of the CT \& FV Methods}
\label{sec:CT-FV-consistency}

To build a numerical scheme based on the CT method for the magnetic
flux, and a FV method for the remaining conserved quantities, it is
very important that the surface and volume averaged magnetic field
components (and their fluxes) be coupled in a consistent manner.  One
common approach \cite{Dai-Woodward-98,RMJF98,Balsara-Spicer} (which
we also follow) is to define the volume-averaged magnetic field
components at cell centers as equal to the average of the
area-averaged values at cell faces, i.e.
\begin{eqnarray}
B_{x,i,j}^n & = &
\frac{1}{2} \left( B_{x,i-1/2,j}^n + B_{x,i+1/2,j}^n \right) 
\label{eq:Bx-CT-FV} \\
B_{y,i,j}^n & = &
\frac{1}{2} \left( B_{y,i,j-1/2}^n + B_{y,i,j+1/2}^n \right)
\label{eq:By-CT-FV}
\end{eqnarray}
which is sufficient for second order accuracy.  However, as shown
below this choice implies a specific relationship between the
numerical fluxes for the induction equation as integrated in the CT
and FV formulations.

\par
The expressions which describe the coupling of the fluxes in the CT \&
FV methods are a direct result of requiring that equations
(\ref{eq:Bx-CT-FV}) and (\ref{eq:By-CT-FV}) also hold at time
$t^{n+1}$.  Subtracting equation (\ref{eq:Bx-CT-FV}) from an
equivalent expression at time $t^{n+1}$, and substituting equations
(\ref{eq:fv_int}) and (\ref{eq:ct_Bx_int}) for the time differences of
the volume- and area-averaged magnetic field respectively we find
\begin{eqnarray}
\hat{e}_{B_x} \cdot 
 \left( F_{y,i,j-1/2}^{n+1/2} - F_{y,i,j+1/2}^{n+1/2} \right)  & = &
\frac{1}{2}
 \left( \E_{z,i-1/2,j-1/2}^{n+1/2} - \E_{z,i-1/2,j+1/2}^{n+1/2} \right) \\
 & + & \frac{1}{2}
 \left( \E_{z,i+1/2,j-1/2}^{n+1/2} - \E_{z,i+1/2,j+1/2}^{n+1/2} \right) ~.
\end{eqnarray}
where $\hat{e}_{B_x}$ is a unit vector for the $B_x$ component of the
flux vector.  It follows that
\beq
\hat{e}_{B_x} \cdot F_{y,i,j \pm 1/2}^{n+1/2} = \frac{1}{2} \left(
 \E_{z,i-1/2,j \pm 1/2}^{n+1/2} + \E_{z,i+1/2,j \pm 1/2}^{n+1/2} \right)
\label{eq:ct_fv_x-consistent}
\eeq
which is consistent with the observation that the flux averages used
in the finite volume method (equation \ref{eq:fv_yflux}) are spatial
averages of the electric fields used in the constrained transport
method (equation \ref{eq:ct_Ez_int}).  Repeating this analysis for
$B_y$ we find
\beq
\hat{e}_{B_y} \cdot F_{x,i \pm 1/2,j}^{n+1/2} = \frac{-1}{2} \left(
 \E_{z,i \pm 1/2,j-1/2}^{n+1/2} + \E_{z,i \pm 1/2,j+1/2}^{n+1/2} \right)
\label{eq:ct_fv_y-consistent}
\eeq
where $\hat{e}_{B_y}$ is a unit vector for the $B_y$ component of the
flux vector.  Functionally, equations (\ref{eq:ct_fv_x-consistent})
and (\ref{eq:ct_fv_y-consistent}) imply that one must replace the
Godunov fluxes for the volume averaged $x$- and $y$-components of the
magnetic field with the average of the corner centered $\E_z$,
regardless of the details of the CT algorithm used to compute the
latter.  Thus, equations (\ref{eq:ct_fv_x-consistent}) and
(\ref{eq:ct_fv_y-consistent}) can be thought of as a corrector step
that makes the predicted Godunov fluxes given by the Riemann solver
consistent with the CT fluxes.  Clearly, the CT algorithm used to
compute the corner centered $\E_z$ will directly impact the accuracy
and stability of the underlying Godunov scheme; in
\S\ref{sec:CT_Algorithms} we address the problem of constructing CT
algorithms and the properties they should possess.

\section{First Order CT Godunov Scheme}

\par
In this section we will construct and test a single step, two
dimensional, first order integration algorithm based upon the
piecewise parabolic method (PPM).  The simplicity of this algorithm
allows us to develop and test two of the most important elements of
this paper: the calculation of the interface states in MHD and the
systematic construction of CT algorithms.  Moreover, the resulting
algorithm (apart from the update step) is essentially the first half
of the CTU + CT integration algorithm described in
{}\S\ref{sec:ctu_ct_Godunov}.

\par
In section {}\ref{sec:lr_states} we describe the calculation of the
interface states in the PPM algorithm for the system of ideal MHD.
This interface state calculation involves a characteristic evolution
of a dimensionally split system.  In this step we will find the
appearance of truly multidimensional terms in the induction equation
which are proportional to $\partial B_x/\partial x$ and $\partial B_y
/ \partial y$.  When these terms are included in the dimensionally
split, linearized system of equations which are used to perform the
characteristic evolution, they take the appearance of ``source
terms''.  We present two \emph{gedanken} experiments which show that
these multidimensional terms are essential to accurately predicting
the time evolution in the interface states.

\par
In \S\ref{sec:CT_Algorithms} we address the question of the
consistency of the CT algorithm with the underlying finite volume
integration algorithm.  Repeating the arguments which lead to
equations (\ref{eq:ct_fv_x-consistent}) and
(\ref{eq:ct_fv_y-consistent}), we present an example of a CT algorithm
which has insufficient dissipation and fails to reduce to the
underlying integration algorithm for plane-parallel, grid-aligned
flows.  We proceed to describe a systematic approach to constructing
a CT algorithm which can be applied with any approximate Riemann
solver and reduces exactly to the underlying finite volume integration
algorithm for plane-parallel, grid-aligned flows.  In
{}\S\ref{sec:ss_ct_Godunov} we present a numerical study comparing
three, surprisingly simple, CT algorithms each of which differs only
in its dissipation for truly multidimensional problems.

\subsection{Calculating the Interface States}
\label{sec:lr_states}

\par
The algorithms presented in this paper are built upon the the
piecewise parabolic method (PPM).  For a thorough discussion of PPM or
its linear variant PLM and its implementation we refer the reader to
the excellent descriptions in
{}\cite{Colella-Woodward-PPM,Saltzman,Miller-Colella}.  Roughly
speaking the PPM algorithm can be broken down into three steps:
spatial reconstruction, characteristic evolution, and flux evaluation.
The purpose of these first two steps is to calculate a one-sided
estimate of the time averaged state at the left- or right-hand sides of
a particular grid cell interface.  With these interface states in
hand, the interface flux may be calculated via either an exact or
approximate Riemann solver.

\par
The calculation of the interface states in PPM is performed in
primitive variables, and is a one-dimensional algorithm.  However, in
the two dimensional $(x,y)$ system of equations for ideal MHD there
appear terms proportional to $\partial B_x/\partial x$ and $\partial
B_y / \partial y$ which are not present in the truly one-dimensional
system.  In primitive variables, these terms only appear in the
induction equation, which in component form is
\beq
\frac{\partial B_x}{\partial t} 
+ \frac{\partial}{\partial y} \left(v_y B_x - B_y v_x \right) = 0
\label{eq:Bx_induction}
\eeq
\beq
\frac{\partial B_y}{\partial t} 
+ \frac{\partial}{\partial x} \left(v_x B_y - B_x v_y \right) = 0
\label{eq:By_induction}
\eeq
\beq
\frac{\partial B_z}{\partial t} 
+ \frac{\partial}{\partial x} \left(v_x B_z - B_x v_z \right) 
+ \frac{\partial}{\partial y} \left(v_y B_z - B_y v_z \right) = 0 ~.
\label{eq:Bz_induction}
\eeq
Using the magnetic charge constraint $({\bf \nabla \cdot B}=0)$ these
terms can be eliminated from equation (\ref{eq:Bz_induction}), giving
\beq
\frac{\partial B_z}{\partial t} 
+ \frac{\partial}{\partial x} \left(v_x B_z \right) 
- B_x \frac{\partial v_z}{\partial x}
+ \frac{\partial}{\partial y} \left(v_y B_z \right)
- B_y \frac{\partial v_z}{\partial y} = 0 ~.
\eeq
However, no such simplification can be made to equations
(\ref{eq:Bx_induction}) and (\ref{eq:By_induction}).  It is natural to
ask just how important these terms are and what role they play in the
evolution of the magnetic field.  A few \emph{gedanken} experiments
quickly show that they are absolutely essential, and at times are the
dominant term in the equation.

\par
Perhaps the most trivial example of a case in which these terms play
an important role is for stationary solutions.  As a concrete example,
consider a circularly polarized Alfv\'en wave oriented at some oblique
angle to the grid.  One may always choose a reference frame in which
the Alfv\'en wave is stationary. Because the wave is oriented oblique
to the grid, $\partial B_x/\partial x$ and $\partial B_y / \partial y$
are non-zero throughout the domain (except at extrema).  In addition,
for a standing Alfv\'en wave, the velocities are of the order of the
Alfv\'en speed.  In this case, the term $v_x (\partial B_y/\partial
y)$ in equation (\ref{eq:Bx_induction}) is not a small term and in
fact, it must exactly balance the remaining term $\partial/\partial
y(v_y B_x) - B_y (\partial v_x/\partial y)$ in order to hold the
stationary solution.  A similar situation also holds for equation
(\ref{eq:By_induction}).

\par
As a second example, consider the simple advection of a magnetic field
loop in the $(x,y)$-plane.  Specifically, let $(\rho,~P,~{\bf v})=$ a
constant with ${\bf v} = v_x {\bf\hat{i}}$, $B_z=0$, and a circular
magnetic field loop in the $(x,y)$-plane of sufficiently weak strength
that $\beta = 2 P/B^2 \gg 1$.  This problem is equivalent to the
advection of a passive scalar, the $z$-component of the magnetic
vector potential.  In this case equations (\ref{eq:Bx_induction}) and
(\ref{eq:By_induction}) are to a very good approximation given by
\beq
\frac{\partial B_x}{\partial t} - v_x \frac{\partial B_y}{\partial y} = 0
\label{eq:Bx_induction-advect}
\eeq
\beq
\frac{\partial B_y}{\partial t} + v_x \frac{\partial B_y }{\partial x} = 0 ~.
\label{eq:By_induction-advect}
\eeq
Hence it is clear that for this particular problem the term
$v_x(\partial B_y/\partial y)$ is not only important, but completely
controls the evolution of the $x$-component of the magnetic field.

\par
We conclude that if the calculation of the interface states for
multidimensional ideal MHD includes a characteristic evolution step,
it is necessary to include the influence of the inherently
multidimensional terms in the induction equation.  Since PPM
reconstruction includes a characteristic evolution step, we have found
the following modifications necessary for multidimensional MHD.

\par
We will restrict the description to a single spatial grid cell index
$i$ and consider the reconstruction process in the $x$-direction.  We
begin by calculating the primitive state vector, $V_i=\{\rho, v_x,
v_y, v_z, B_x, B_y, B_z, P\}_i$ and $\tilde{V}_i$ such that
$V_i=(\tilde{V}_i,~B_{x,i})$, associated with $q_i$, the vector of the
cell averaged conserved variables.  Next we apply the PPM algorithm to
calculate the interface states of $\tilde{V}_i$ where the
characteristic evolution step is calculated by solving
\begin{equation}
\frac{\partial \tilde{V}}{\partial t} 
 + A \frac{\partial \tilde{V}}{\partial x}
 = \sigma
\label{eq:wave-prop-src}
\end{equation}
where
\beq
\tilde{V} = \left (
\begin{array}{c}
\rho \\
v_x \\
v_y \\
v_z \\
B_y \\
B_z \\
P
\end{array}
\right ) ~, ~~
\sigma = \left (
\begin{array}{c}
0 \\
0 \\
0 \\
0 \\
v_y (\partial B_x / \partial x)\\
0 \\
0
\end{array}
\right ) ~,
\label{eq:wave-prop-src-terms}
\eeq
\beq
A = \left (
\begin{array}{ccccccc}
v_x & \rho     &  0   &  0   &  0        &  0        & 0      \\
0   &  v_x     &  0   &  0   &  B_y/\rho &  B_z/\rho & 1/\rho \\
0   &   0      &  v_x &  0   & -B_x/\rho &  0        & 0      \\
0   &   0      &  0   &  v_x &  0        & -B_x/\rho & 0      \\
0   &  B_y     & -B_x &  0   &  v_x      & 0         & 0      \\
0   &  B_z     &  0   & -B_x &  0        &  v_x      & 0      \\
0   & \gamma P &  0   &  0   &  0        &  0        & v_x    \\
\end{array}
\right ) ~.
\label{eq:wave-prop-matrix}
\eeq
The matrix $A$ is linearized about the state $V_i$ and the source term
$\sigma$ is taken to be a constant with the only non-zero term
evaluated as $v_{y,i} (B_{x,i+1/2}-B_{x,i-1/2})/\delta x$.  Note that
equation (\ref{eq:wave-prop-src}) includes all of the terms from
equation (\ref{eq:By_induction}).  Denote the interface states
calculated in this procedure as $\tilde{V}^{L}_{i+1/2}$ and
$\tilde{V}^{R}_{i-1/2}$ where the superscripts $(L,~R)$ denote the
left or right hand side of the interface to which they are adjacent.
The final step is to define the primitive states $V^{L}_{i+1/2} =
(\tilde{V}^{L}_{i+1/2}, B_{x,i+1/2})$ and $V^{R}_{i-1/2} =
(\tilde{V}^{R}_{i-1/2}, B_{x,i-1/2})$.  Note a further significant
advantage of using face-centered (staggered) fields: the interface
states of the longitudinal component of the magnetic field do not need
to be reconstructed, and therefore will be continuous.  Moreover,
since monotonicity constraints associated with reconstruction are not
applied, extrema in the longitudinal component of B at interfaces will
be preserved.

\par
The calculation of the $y$-interface states follows this same
procedure, with the matrix $A$ replaced by the equivalent
one-dimensional wave matrix for the $y$-direction and the source term
$\sigma$ containing a non-zero entry for $B_x$ equal to $v_x (\partial
B_y/\partial y)$.

\subsection{Constrained Transport Algorithms}
\label{sec:CT_Algorithms}

\par
In this section we take up the problem of constructing CT algorithms.
In order to identify the properties of a suitable CT algorithm, it is
particularly interesting to consider the limiting case of plane-parallel,
grid-aligned flows.  In this limit $\partial B_x/\partial x = -\partial
B_y / \partial y = 0$ so that there is no longer a difference between area
and volume averaged magnetic fields.  Moreover, if for example $\partial /
\partial x = 0$ then the correct solution to the CT algorithm is readily
obtained via symmetry, e.g. $\E_{z,i+1/2,j+1/2} = \left( \E_{z,i,j+1/2}
+ \E_{z,i+1,j+1/2}\right)/2$.  When a CT algorithm reduces to this,
or an equivalent expression, for plane-parallel grid-aligned flows
we describe it as being consistent with the underlying integration
algorithm, since in this case it will give the identical solution
as the underlying integration algorithm applied to the equivalent
one-dimensional problem.  Furthermore, we seek to construct CT
algorithms which are compatible with any approximate Riemann solver,
e.g. \cite{Cargo-Gallice,Dai-Woodward-97,Einfeldt,S-Li-HLLC}.  Hence,
they should only depend upon the electric field in the flux vector,
not on the structure of the waves which result from the solution of the
Riemann problem.

\subsubsection{Arithmetic Averaging}

\par
Perhaps the simplest, and most often suggested, CT algorithm is based
upon averaging the face centered electric fields obtained from the
underlying integration algorithm, i.e. choose $\E_{z,i+1/2,j+1/2} =
\bar{\E}_{z,i+1/2,j+1/2}$ where
\beq
\bar{\E}_{z,i+1/2,j+1/2} = \frac{1}{4} \left(
\E_{z,i+1/2,j} + \E_{z,i+1/2,j+1} + \E_{z,i,j+1/2} + \E_{z,i+1,j+1/2}
\right) ~.
\label{eq:Ez-avg}
\eeq
Unfortunately, this CT algorithm is not consistent with the underlying
integration algorithm for plane-parallel, grid-aligned flows.  This
behavior is most easily understood when the underlying finite volume
integration algorithm is unsplit.

\par
Consider a plane-parallel, grid-aligned flow in which
$\partial/\partial x=0$.  It follows that
$\E_{z,i,j+1/2}=\E_{z,i+1,j+1/2}$ and $\E_{z,i+1/2,j}=\E_{z,i,j}$.
Inserting these expressions into equation (\ref{eq:Ez-avg}) we find
\beq
\bar{\E}_{z,i+1/2,j+1/2} = \frac{1}{4} \left( \E_{z,i,j} + \E_{z,i,j+1} \right)
+ \frac{1}{2} \E_{z,i,j+1/2} ~.
\label{eq:Ez-avg-PPGAF}
\eeq
Contrast this with the correct solution, which by the assumption of
planar symmetry is simply $\E_{z,i+1/2,j+1/2} = \E_{z,i,j+1/2}$.  In
order to assess the impact this CT algorithm has on the integration
algorithm as a whole, let the FV numerical flux $F_{y,i,j+1/2}$ be
written as
\beq
F_{y,i,j+1/2} = \frac{1}{2}
\left( f_y(q_{i,j}) + f_y(q_{i,j+1}) + D_{i,j+1/2} (q_{i,j} - q_{i,j+1})
\right) ~,
\eeq
where $D_{i,j+1/2}$ is the viscosity-matrix \cite{HLL,Einfeldt}.
Contracting this expression with a unit vector $\hat{e}_{B_x}$ to
extract the $y$-flux of $B_x$ (remembering that $f_y(B_x)=\E_z$) we
have the FV numerical electric field
\beq
\E_{z,i,j+1/2} = \frac{1}{2}(\E_{z,i,j} + \E_{z,i,j+1})
+ \frac{1}{2} \hat{e}_{B_x} D_{i,j+1/2} (q_{i,j} - q_{i,j+1}) ~.
\label{eq:numerical_flux_Ez}
\eeq
Recall that, as discussed in \S\ref{sec:CT-FV-consistency}, for FV +
CT schemes this electric field is essentially a predictor value.  To
obtain the corrector value we begin by inserting the FV numerical
electric field in equation (\ref{eq:numerical_flux_Ez}) into the
$\bar{\E}$ CT algorithm in equation (\ref{eq:Ez-avg-PPGAF}) giving
\beq
\bar{\E}_{z,i+1/2,j+1/2} = \frac{1}{2}(\E_{z,i,j} + \E_{z,i,j+1})
+ \frac{1}{4} \hat{e}_{B_x} D_{i,j+1/2} (q_{i,j} - q_{i,j+1}) ~.
\label{eq:half_visc_flux}
\eeq
Note also that by symmetry we have
$\bar{\E}_{z,i-1/2,j+1/2}=\bar{\E}_{z,i+1/2,j+1/2}$.  Applying
equation (\ref{eq:ct_fv_x-consistent}) we obtain the corrector value
of the FV numerical electric field $\bar{\E}_{z,i,j+1/2} =
\bar{\E}_{z,i+1/2,j+1/2}$ given by equation (\ref{eq:half_visc_flux}).
Comparing equations (\ref{eq:numerical_flux_Ez}) and
(\ref{eq:half_visc_flux}) we find that the numerical viscosity is
reduced by a factor of 2.  Hence it's clear that with the arithmetic
average CT algorithm $\bar{\E}$, the solution algorithm does not
reduce to the underlying integration algorithm for plane-parallel,
grid-aligned flows and the stability of this approach is questionable.
The failure of this simple procedure to reduce to the underlying
integration algorithm for plane-parallel, grid-aligned flows can be
traced back to the lack of a directional bias in the averaging
formula.

\par
The arithmetic average CT algorithm formed the basis of an algorithm
proposed by Balsara and Spicer \cite{Balsara-Spicer}. The need for the
CT algorithm to have a directional biasing was well understood by
these authors.  In their paper, they presented two switches which
serve as local, multidimensional sensors for magnetosonic shocks.  The
authors then applied weighting coefficients which impart a directional
bias to the CT algorithm.  

\par
However, the recognition that in equation (\ref{eq:half_visc_flux})
the viscous flux contribution to the corner value for $\E_z$ in the CT
algorithm is simply too small by a factor of two, suggests that by
doubling it we could recover the proper directional biasing.  To that
end we define
\begin{eqnarray}
\hat{\E}_{z,i+1/2,j+1/2} & \equiv & 2 \bar{\E}_{z,i+1/2,j+1/2} 
\nonumber \\
& - & \frac{1}{4} 
\left( \E_{z,i,j} + \E_{z,i,j+1} + \E_{z,i+1,j} + \E_{z,i+1,j+1} \right)
\end{eqnarray}
which can be written out explicitly as
\begin{eqnarray}
\hat{\E}_{z,i+1/2,j+1/2} & \equiv & \frac{1}{2} 
\left( \E_{z,i+1/2,j} + \E_{z,i+1/2,j+1} + \E_{z,i,j+1/2} + \E_{z,i+1,j+1/2}
\right) \nonumber \\
& - & \frac{1}{4}
\left( \E_{z,i,j} + \E_{z,i,j+1} + \E_{z,i+1,j} + \E_{z,i+1,j+1} \right) ~.
\label{eq:NoDiss_CT_Ez}
\end{eqnarray}
Repeating the arguments leading to equation (\ref{eq:Ez-avg-PPGAF})
using the CT algorithm defined by equation (\ref{eq:NoDiss_CT_Ez}) one
finds that the resulting scheme reduces to the underlying finite
volume integration algorithm for plane-parallel, grid-aligned flows.
It is not clear, however, from the ad-hoc construction described here
how well such an algorithm will behave for truly multidimensional
flows.

\par
In the following section we describe a systematic approach to
constructing a CT algorithm which by design reverts to the underlying
finite volume integration algorithm for plane-parallel, grid-aligned
flows.  Two entirely new CT algorithms will be constructed based upon
different approximations.  We will also find that the CT algorithm
described by equation (\ref{eq:NoDiss_CT_Ez}) can be understood as a
limiting case of one of the CT algorithms constructed in the next
section.  Despite the simplicity of the CT algorithms constructed in
the following section, we will see in {}\S\ref{sec:ss_ct_Godunov} that
they behave surprisingly well on multidimensional tests.

\subsubsection{Systematic Construction of CT Algorithms}
\label{sec:build_CT}

\par
The approach described here is based upon the observation that a CT
algorithm can be thought of as the inverse of the consistency
relations given by equations (\ref{eq:ct_fv_x-consistent}) and
(\ref{eq:ct_fv_y-consistent}).  In this sense, the CT algorithm can be
thought of as a reconstruction, or integration procedure.  Note that
the interface fluxes which are calculated as part of the base
integration algorithm are midpoint values, centered spatially on the
grid cell face, and averaged temporally over the time step.  This
suggests that we consider the CT algorithm, which calculates a time
averaged value of $\E_z$ at the grid cell corner, to be a spatial
integration procedure.  For example, given a face centered value
$\E_{z,i+1/2,j}$ we seek an estimate of $(\partial \E_z/\partial
y)_{i+1/2,j+1/4}$ giving one value for
\beq
\E_{z,i+1/2,j+1/2} = \E_{z,i+1/2,j} + \frac{\delta y}{2} 
\left(\frac{\partial \E_z}{\partial y}\right)_{i+1/2,j+1/4} ~.
\eeq
Clearly in two dimensions one may integrate from any one of the four
nearest face centers to the corner and generally the resulting values
for $\E_{z,i+1/2,j+1/2}$ will differ.  In the CT algorithms presented
here, we will use the arithmetic average of these four values giving
\begin{eqnarray}
\E_{z,i+1/2,j+1/2} & = &
\frac{1}{4} \left(
\E_{z,i+1/2,j} + \E_{z,i+1/2,j+1} + \E_{z,i,j+1/2} + \E_{z,i+1,j+1/2} 
\right) \nonumber \\
& + & \frac{\delta y}{8} \left(
\left(\frac{\partial \E_z}{\partial y}\right)_{i+1/2,j+1/4} - 
\left(\frac{\partial \E_z}{\partial y}\right)_{i+1/2,j+3/4}
\right) \nonumber \\
& + & \frac{\delta x}{8} \left(
\left(\frac{\partial \E_z}{\partial x}\right)_{i+1/4,j+1/2} - 
\left(\frac{\partial \E_z}{\partial x}\right)_{i+3/4,j+1/2}
\right) ~.
\label{eq:2d_CT_Ez}
\end{eqnarray}
The construction of this CT algorithm is completed by specifying a way
to calculate the derivatives of $\E_z$ on the grid cell face.

\par
To calculate $(\partial \E_z/\partial x)$ and $(\partial \E_z/\partial
y)$ at grid cell faces, we propose to use an approximate solution for
the evolution equations for $(\partial B_x/\partial x)$ and $(\partial
B_y/\partial y)$.  At a $y$-interface we differentiate the induction
equation for $B_x$ giving
\beq
\frac{\partial}{\partial t} \left(\frac{\partial B_x}{\partial x} \right) + 
\frac{\partial}{\partial y} \left(\frac{\partial \E_z}{\partial x}\right) 
= 0 ~.
\eeq
Similarly, at an $x$-interface we differentiate the induction equation
for $B_y$ giving
\beq
\frac{\partial}{\partial t} \left(\frac{\partial B_y}{\partial y} \right) - 
\frac{\partial}{\partial x} \left(\frac{\partial \E_z}{\partial y}\right) 
= 0 ~.
\eeq
Since these expression are still in conservation form it suggests that
we may calculate an interface value for $(\partial \E_z/\partial x)$
at $y$-interfaces and $(\partial \E_z/\partial y)$ at $x$-interfaces
using for example an HLL or Lax-Friedrichs flux.  To evaluate these
fluxes, we need estimates for the gradients of $(\partial \E_z
/\partial x)$, $(\partial \E_z/\partial y)$, $(\partial B_x/\partial
x)$ and $(\partial B_y/\partial y)$ on either side of the interface.

\par
For the single step, CT Godunov algorithm which we are considering in
this section, we calculate these derivatives as follows.  For
$(\partial B_x/\partial x)$ we difference the interface and cell
center values giving
\beq
\left(\frac{\partial B_x}{\partial x} \right)_{i+1/4,j} = 
\frac{2}{\delta x} \left( B_{x,i+1/2,j} - B_{x,i,j}   \right) ~.
\eeq
For $(\partial \E_z/\partial x)$ we difference the face centered
$\E_{z,i+1/2,j}$ which comes directly from the Riemann solver and the
cell center value $\E_{z,i,j}$ evaluated in the cell center state
$q_{i,j}^n$ giving
\beq
\left(\frac{\partial \E_z}{\partial x} \right)_{i+1/4,j} = 
\frac{2}{\delta x} \left( \E_{z,i+1/2,j} - \E_{z,i,j} \right) ~.
\label{eq:Ez_gradient}
\eeq
The values for $(\partial B_y/\partial y)$ and $(\partial \E_z
/\partial y)$ are given by analogous expressions.

\par
Pursuing the Lax-Friedrichs estimate with a maximum wave speed
$\alpha$ we find
\begin{eqnarray}
\left(\frac{\partial \E_z}{\partial x}\right)_{i+1/4,j+1/2} & = &
\frac{1}{\delta x}
\left(\E_{z,i+1/2,j} - \E_{z,i,j} + \E_{z,i+1/2,j+1} - \E_{z,i,j+1} \right)
 \nonumber \\
& + & \frac{\alpha}{\delta x} \left( B_{x,i+1/2,j} - B_{x,i,j} 
- B_{x,i+1/2,j+1} + B_{x,i,j+1} \right) 
\end{eqnarray}
and 
\begin{eqnarray}
\left(\frac{\partial \E_z}{\partial y}\right)_{i+1/2, j+1/4} & = &
\frac{1}{\delta y}
\left(\E_{z,i,j+1/2} - \E_{z,i,j} + \E_{z,i+1,j+1/2} - \E_{z,i+1,j} \right)
 \nonumber \\
& + & \frac{\alpha}{\delta y} \left( B_{y,i+1,j+1/2} - B_{y,i+1,j} 
- B_{y,i,j+1/2} + B_{y,i,j} \right) ~.
\end{eqnarray}
Repeating this procedure for the two remaining gradients and
inserting the results into equation (\ref{eq:2d_CT_Ez}) we obtain
\begin{eqnarray}
\E^\alpha_{z,i+1/2, j+1/2} & = &
\frac{1}{2} \left(\E_{z,i,j+1/2} + \E_{z,i+1,j+1/2} + 
\E_{z,i+1/2,j} + \E_{z,i+1/2,j+1} \right) \nonumber \\
& - & \frac{1}{4} \left(\E_{z,i,j} + \E_{z,i+1,j}
+ \E_{z,i,j+1} + \E_{z,i+1,j+1} \right) \nonumber \\
& + & \frac{\alpha}{8} \left( B_{x,i+1/2,j} - B_{x,i,j} 
- B_{x,i+1/2,j+1} + B_{x,i,j+1} \right) \nonumber \\
& + & \frac{\alpha}{8} \left( B_{x,i+1/2,j} - B_{x,i+1,j} 
- B_{x,i+1/2,j+1} + B_{x,i+1,j+1} \right) \nonumber \\
& + & \frac{\alpha}{8} \left( B_{y,i+1,j+1/2} - B_{y,i+1,j} 
- B_{y,i,j+1/2} + B_{y,i,j} \right) \nonumber \\
& + & \frac{\alpha}{8} \left( B_{y,i+1,j+1/2} - B_{y,i+1,j+1} 
- B_{y,i,j+1/2} + B_{y,i,j+1} \right) ~.
\label{eq:LF_CT_Ez}
\end{eqnarray}
One may readily show that for plane-parallel, grid-aligned flows, this
CT algorithm will properly recover the associated one-dimensional
solution for the underlying integration algorithm. Hereafter, we refer
to equation (\ref{eq:LF_CT_Ez}) as the $\E^\alpha_z$ CT algorithm.

\par
It is particularly interesting to note that the $\alpha=0$ limit of
equation (\ref{eq:LF_CT_Ez}) gives equation (\ref{eq:NoDiss_CT_Ez}).
Hence we may now understand equation (\ref{eq:NoDiss_CT_Ez}) as being
equivalent to the integration and averaging procedure described here
with the approximation
\beq
\left(\frac{\partial \E_z}{\partial y}\right)_{i+1/2,j+1/4} = 
\frac{1}{2}\left(\frac{\partial \E_z}{\partial y} \right)_{i,j+1/4} + 
\frac{1}{2}\left(\frac{\partial \E_z}{\partial y} \right)_{i+1,j+1/4} ~.
\eeq
Clearly this is not an upwinded approximation, suggesting that we
should find some level of oscillations present in using this CT
algorithm for multidimensional flows.  However, since the dissipation
arising from the terms proportional to $\alpha$ in the $\E^\alpha_z$
CT algorithm are only important for truly multidimensional flows, it
is not clear if their neglect will have a substantive impact on the
first order integration algorithm which we are considering here.  For
that reason, henceforth we will refer to the $\alpha=0$ limit of the
$\E^\alpha_z$ CT algorithm as $\E^\circ_z$, and include it in the
tests in the following section.

\par
We now have two CT algorithms: the $\E^\circ_z$ algorithm given by
equation (\ref{eq:NoDiss_CT_Ez}), and the $\E^\alpha_z$ algorithm
given by equation (\ref{eq:LF_CT_Ez}).  As our final CT algorithm, we
note that for the special case of advection, $(\partial \E_z/\partial
y)$ at an $x$-interface should be selected in an upwind fashion
according to the contact mode.  As such, we suggest that upwinding
$(\partial \E_z/\partial y)$ at $x$-interfaces (and similarly
$(\partial \E_z/\partial x)$ at $y$-interfaces) according to the
contact mode may be sufficient to lead to a stable, non-oscillatory
integration algorithm.  Specifically, we choose
\beq
\left(\frac{\partial \E_z}{\partial y}\right)_{i+1/2,j+1/4} = \left \{
\begin{array}{ll}
(\partial \E_z/\partial y)_{i,j+1/4} & \textrm{for}~ v_{x,i+1/2,j} > 0 \\
(\partial \E_z/\partial y)_{i+1,j+1/4} & \textrm{for}~ v_{x,i+1/2,j} < 0 \\
\frac{1}{2}\left(
(\partial \E_z/\partial y)_{i,j+1/4} + (\partial \E_z/\partial y)_{i+1,j+1/4}
\right) & \textrm{otherwise} ~.
\end{array} 
\right .
\label{eq:Ndup_CT_Ez}
\eeq
Note that this simply depends upon the sign of the mass flux, not the
details of the solution of the Riemann problem at the interface and
therefore can be applied with any approximate or exact Riemann solver.
An analogous expression holds for the remaining three interface
gradients of $\E_z$.  We will refer to the CT algorithm which results
from combining this approximation for the gradients of $\E_z$ with
equation (\ref{eq:2d_CT_Ez}) as the $\E^c_z$ CT algorithm.  By design
this CT algorithm reduces to the underlying integration algorithm for
plane-parallel grid-aligned flows and is properly upwinded in a
multidimensional sense for the simple case of magnetic field
advection.

\par
In this section we've presented a simple approach to constructing a CT
algorithm which reduces exactly to the base integration algorithm for
plane-parallel, grid-aligned flows.  By design, the CT algorithms
constructed here differ only in their numerical viscosity for
multidimensional problems.  It should be noted that the approach
presented here can readily be incorporated into other numerical
schemes, such as wave propagation algorithms
{}\cite{Langseth-LeVeque}.  The CT algorithms described here can also
be applied to integration algorithms based upon spatial reconstruction
to the grid cell corners as well {}\cite{Balsara-04}.  While this
might seem surprising following this presentation, note that the
factors $(\delta x, \delta y)$ cancel in the $(\E^\alpha_z,
\E^\circ_z, \E^c_z)$ CT algorithms.  For such integration algorithms,
the cell center value $\E_{z,i,j}$ should be replaced with the value
of $\E_z$ calculated in the reconstructed fluid state at the grid cell
corner, e.g. $q_{i+1/2,j+1/2}$.  With this choice, the CT algorithms
presented here will reduce to the base integration algorithms for
plane-parallel, grid-aligned flows.

\par
We note that another CT algorithm with the properties that it reduces
to the base integration algorithm for plane-parallel, grid-aligned
flows has recently been presented and tested elsewhere
{}\cite{Londrillo-DelZanna04}.  In particular, the authors of that
paper present a general framework for combining CT and Godunov-type
schemes and two specific implementations for their positive and
central-type schemes.  A direct comparison between their approach and
ours is somewhat complex in the general case.  In the specific case of
a first order Godunov scheme, one can show (using equations 41 - 47 in
{}\cite{Londrillo-DelZanna04}) that their CT algorithm is identical to
the $\E^\circ_z$ CT algorithm constructed here, although this is not
immediately obvious from the description of their framework.  For the
more complex CT algorithms developed in our paper, it is likely that
they too can be cast in the framework described by Londrillo \& Del
Zanna {}\cite{Londrillo-DelZanna04}, although we have not attempted to
do so.

\subsection{First Order CT Godunov Tests}
\label{sec:ss_ct_Godunov}

\par
The integration algorithm utilized in this section is easily assembled
from the elements described in the preceding sections.  Starting at
time $t^n$ we calculate the $x$- and $y$-interface states as described
in section \ref{sec:lr_states} using the PPM algorithm.  Next we use
an approximate Riemann solver to calculate a flux at each grid cell
interface.  In the tests presented here we use a Roe linearization
\cite{Cargo-Gallice}.  Finally, we apply one of the three
$(\E^\alpha_z, \E^\circ_z, \E^c_z)$ CT algorithms.  The resulting
integration algorithm is first order accurate and subject to a
restrictive CFL stability limit.  In the tests presented in this
section we use a time step
\beq
\delta t = 0.4 \min \left(
\frac{\delta x}{|\lambda^{max}_x|},
\frac{\delta y}{|\lambda^{max}_y|} \right) 
\eeq
where $\lambda^{max}_{x,y}$ indicates the fastest wave mode speed in
the $x$- or $y$-direction.

\subsubsection{Field Loop Advection}
\label{sec:ss_ct_Godunov:field_loop}

The first problem we consider is the advection of a weak magnetic
field loop.  The computational domain extends from $-1\le x \le 1$,
and $-0.5 \le y \le 0.5$, is resolved on a $2N \times N$ grid and has
periodic boundary conditions on both $x$- and $y$-boundaries.  For the
tests presented in this section we take $N=64$.  The mass density
$\rho=1$ and the gas pressure $P=1$.  The velocity components $v_x =
v_0 \cos(\theta)$, $v_y = v_0 \sin(\theta)$, $v_z=0$ where 
$\cos(\theta)=2/\sqrt{5}$ and $\sin(\theta)=1/\sqrt{5}$.  In the
diffusion tests we set $v_0=0$ while in the advection tests we set
$v_0=\sqrt{5}$ so that by $t=1$ the field loop will have been advected
around the grid one complete orbit along the grid diagonal.  The
$z$-component of the magnetic field $B_z=0$ while the in plane
components $B_x$ and $B_y$ are initialized from the $z$-component of
the magnetic vector potential where
\beq
A_z \equiv \left \{
\begin{array}{ll}
A_0 (R - r) & \textrm{for}~ r \le R \\
0           & \textrm{for}~ r  >  R
\end{array}
\right .
\eeq
where $A_0=10^{-3}$, $R=0.3$ and $r=\sqrt{x^2+y^2}$.  Thus for $r \le
R$, $\beta = 2 P/B^2 = 2\times 10^{6}$ and the magnetic field is
essentially a passive scalar.

\par
\begin{figure}[htb]
\begin{center}
\includegraphics*[width=2.5 in]{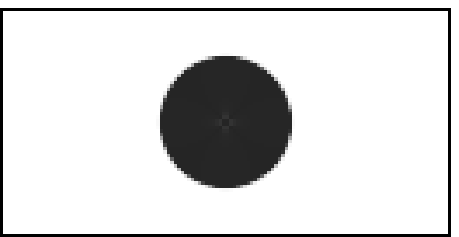} \hfill
\includegraphics*[width=2.5 in]{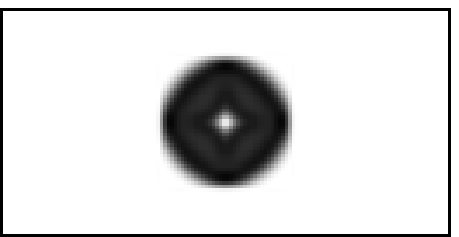} \\
\includegraphics*[width=2.5 in]{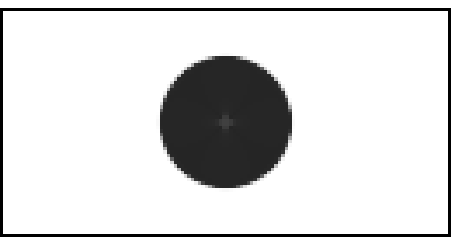} \hfill
\includegraphics*[width=2.5 in]{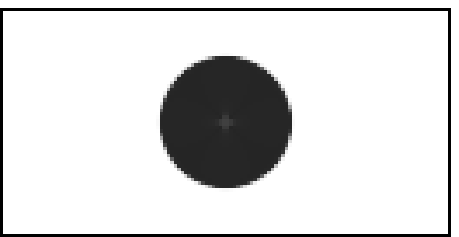}
\end{center}
\caption{Grey-scale images of the magnetic pressure $(B_x^2 + B_y^2)$ at 
$t=0$ (top left) and at $t=2$ for a stationary medium ($v_0=0$) using
the $\E^\alpha_z$ (top right), the $\E^\circ_z$ (bottom left) and the
$\E^c_z$ (bottom right) CT algorithm.}
\label{fig:loop-diffuse}
\end{figure}

In the first test we consider the diffusion of the field loop.  In
figure \ref{fig:loop-diffuse} we compare grey-scale images of the
magnetic pressure $(B_x^2 + B_y^2)$ at $t=0$ to the evolved results at
$t=2$ using the three $(\E^\alpha_z, \E^\circ_z, \E^c_z)$ CT
algorithms.  Clearly the $\E^\alpha_z$ CT algorithm leads to an
unacceptable amount of diffusion compared to the other two, as
evidenced by the emergence of a hole at the center caused by
reconnection.  The $(\E^\circ_z, \E^c_z)$ CT algorithms lead to
essentially identical results and a very small amount of diffusion.
Apparently the additional dissipation included in the $\E^\alpha_z$ CT
algorithm is not necessary for stability in this test.

\par
\begin{figure}[bht]
\begin{center}
\includegraphics*[width=2.5 in]{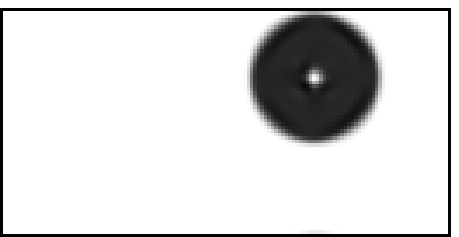}  \hfill 
\includegraphics*[width=2.5 in]{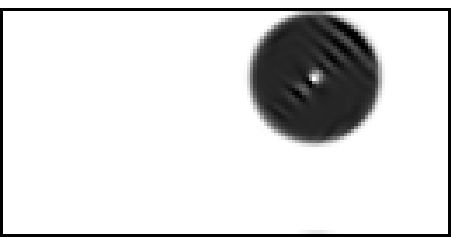} \hfill
\includegraphics*[width=2.5 in]{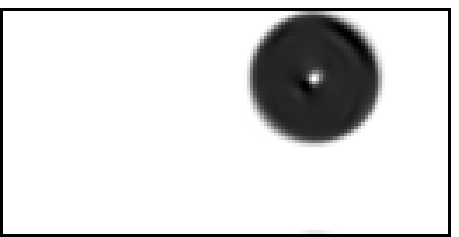}
\end{center}
\caption{Grey-scale images of the magnetic pressure $(B_x^2 + B_y^2)$
at $t=0.19$ for an advected field loop ($v_0=\sqrt{5}$) using the
$\E^\alpha_z$ (top left), $\E^\circ_z$ (top right) and $\E^c_z$ (bottom) CT
algorithm.}
\label{fig:loop-advect-early}
\end{figure}
\begin{figure}[bht]
\begin{center}
\includegraphics*[width=2.5 in]{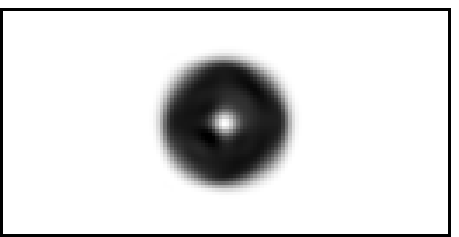}  \hfill 
\includegraphics*[width=2.5 in]{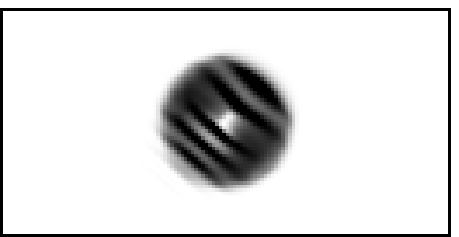} \hfill
\includegraphics*[width=2.5 in]{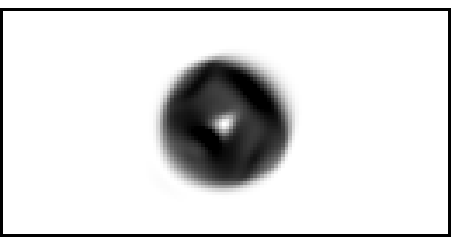}
\end{center}
\caption{Grey-scale images of the magnetic pressure $(B_x^2 + B_y^2)$
at $t=2$ for an advected field loop ($v_0=\sqrt{5}$) using the
$\E^\alpha_z$ (top left), $\E^\circ_z$ (top right) and $\E^c_z$ (bottom) CT
algorithm.}
\label{fig:loop-advect-late}
\end{figure}

Next we consider the advection of the magnetic field loop.  Unlike in
the stationary field loop test, here the evolved results appear quite
similar for the $\E^\alpha_z$ and $\E^c_z$ CT algorithms.  In figure
{}\ref{fig:loop-advect-early} we present grey-scale images of the
magnetic pressure at $t=0.19$ for the three $(\E^\alpha_z, \E^\circ_z,
\E^c_z)$ CT algorithms.  At this time, the $\E^\alpha_z$ and $\E^c_z$
CT algorithms give quite similar results.  In contrast, the
$\E^\circ_z$ CT algorithm appears to have insufficient dissipation
leading to an oscillatory solution.  These observations are consistent
with the comments in \S\ref{sec:build_CT} regarding the upwinding of
the gradients of $\E_z$ at the interfaces.  In figure
{}\ref{fig:loop-advect-late} we present grey-scale images of the
magnetic pressure at $t=2$ for the three $(\E^\alpha_z, \E^\circ_z,
\E^c_z)$ CT algorithms.  By this time the oscillations present using the 
$\E^\circ_z$ CT algorithm have come to dominate the solution.  The
results from the $\E^\alpha_z$ and $\E^c_z$ CT algorithms continue to
remain quite similar, implying that the dissipation of the first order
scheme is comparable to the dissipation in the $\E^\alpha_z$ CT
algorithm.  Note also the similarity to the magnetic pressure image in
figure \ref{fig:loop-diffuse} for the $\E^\alpha_z$ CT algorithm.


\par
From the results of these tests we are led to conclude that of the
three algorithms, the $\E^c_z$ CT algorithm is preferable.  The
$\E^\alpha_z$ CT algorithm leads to stable, yet diffusive results for
stationary problems.  The $\E^\circ_z$ CT algorithm appears to have
insufficient dissipation for advection problems leading to oscillatory
results.  One might naturally wonder, however, if these results are
biased to favor the $\E^c_z$ CT algorithm by design.  In the following
section we present additional tests of these CT algorithms where wave
modes other than the contact mode play an important role in the
solution.  We note in passing that the source terms described in
{}\S\ref{sec:lr_states} are absolutely essential to obtain the results
presented here.  If they had been omitted, the field loop
disintegrates in oscillations before completing a fraction of an
orbital period.

\subsubsection{Circularly Polarized Alfv\'en Wave}
\label{sec:ss_ct_Godunov:cp_alfven}

In a recent paper T\'oth \cite{Toth-divB} described a test problem
involving the evolution of traveling and standing circularly polarized
Alfv\'en waves in a periodic domain.  This test problem is interesting
from the point of view that the initial conditions are nonlinear
solutions to the equations of ideal MHD.  Unfortunately, their
efficacy as a discriminating test for multidimensional MHD codes has
been hindered slightly {}\cite{Pen-MHD-03,Londrillo-DelZanna04} by the
fact that they are susceptible to a parametric instability
{}\cite{Goldstein,DelZanna-ea-01}.  Nevertheless, we have found this
to be a useful test and find no indication of instability for the
parameters adopted here.

\par
The initial conditions we utilize here are slightly different than in
the original description \cite{Toth-divB}.  The computational domain
extends from $0\le x \le \sqrt{5}$, and $0 \le y \le \sqrt{5}/2$, is
resolved on a $2N \times N$ grid and has periodic boundary conditions
on both $x$- and $y$-boundaries.  For the tests presented in this
section we take $N=8$. The Alfv\'en wave propagates at an angle
$\theta = \tan^{-1}(2) \approx 63.4^\circ$ with respect to the
$x$-axis and has a wavelength $\lambda=1$.  The mass density $\rho=1$
and the gas pressure $P=0.1$.  The velocity and magnetic field
components are most easily described in a rotated coordinate system
\begin{eqnarray}
x_1 & = &  x \cos \theta + y \sin \theta \label{eq:coord_trans_x} \\
x_2 & = & -x \sin \theta + y \cos \theta \label{eq:coord_trans_y} \\
x_3 & = & z \label{eq:coord_trans_z}
\end{eqnarray}
such that the Alfv\'en wave propagates along the $x_1$ axis.  The
magnetic field components $B_1 = 1$, $B_2 = 0.1\sin(2\pi x_1)$, and
$B_3 = 0.1\cos(2\pi x_1)$.  The velocity components $v_1 = (0,1)$ for
traveling or standing Alfv\'en waves respectively, $v_2 = 0.1\sin(2\pi
x_1)$, and $v_3 = 0.1\cos(2\pi x_1)$.  With this set of initial
conditions and $v_1=0$ the Alfv\'en wave will travel a
distance of one wavelength $\lambda$ in a time $t=1$.

\par
\begin{figure}[htb]
\begin{center}
\includegraphics*[width=2.5in]{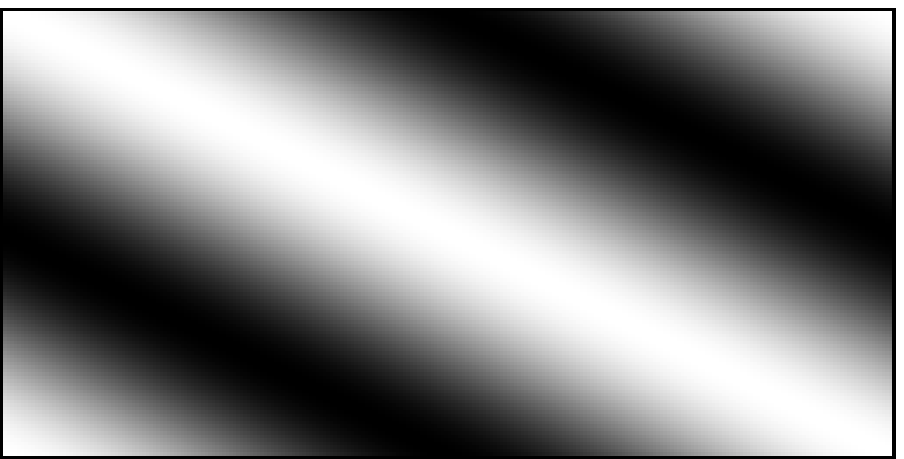} \hfill
\includegraphics*[width=2.5in]{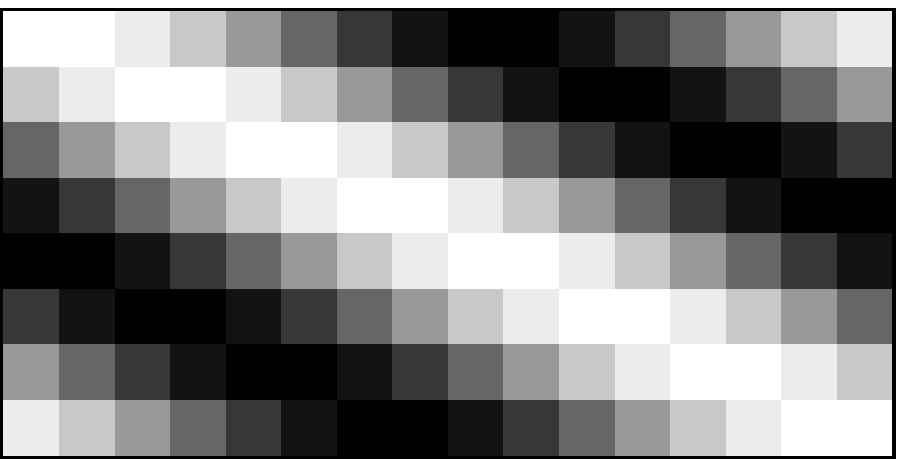}
\end{center}
\caption{Plot of $B_z$ at the initial time for the circularly 
polarized Alfv\'en wave problem at high resolution, $N=128$, (left) 
and at the resolution plotted in figure \ref{fig:cpaw-1step} (right).}
\label{fig:cpaw-1step-Bz}
\end{figure}

To better illustrate the geometry of this problem, a high resolution
image of the out of plane component of the magnetic field, $B_z$ is
presented in figure {}\ref{fig:cpaw-1step-Bz}.  Also included in this
figure is an image of $B_z$ for the resolution tested $(N=8)$ in order
to emphasize that the coarsest resolution for this wave is in the
$y$-direction (with only 8 grid points per wavelength), and that there
are essentially two complete wavelengths across the grid diagonal.  We
have found that low resolution tests such as presented below are much
more informative, since differences between algorithms are generally
largest in this case.  Note that, just as in the field loop problem,
the in-plane components of the magnetic field $(B_x,B_y)$ are
initialized via the $z$-component of the appropriate magnetic vector
potential.

\par
\begin{figure}[htb]
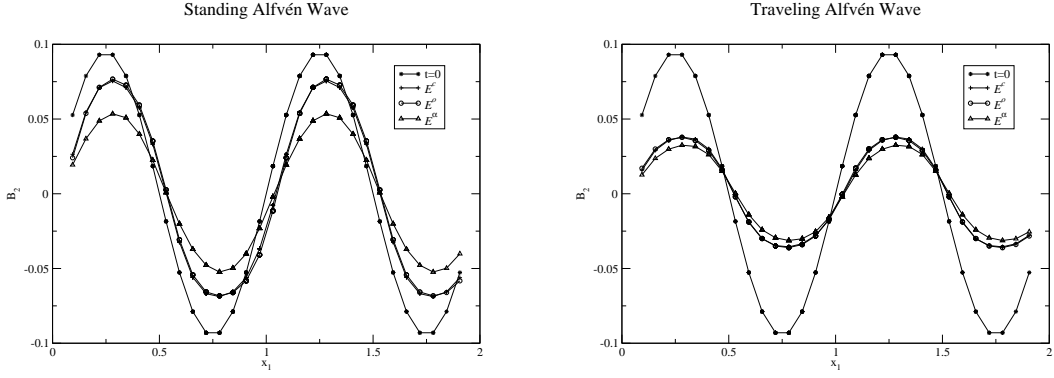

\begin{center}
\includegraphics*[width=2.5in]{standing.cpaw.16x8.eps} \hfill
\includegraphics*[width=2.5in]{traveling.cpaw.16x8.eps} 
\end{center}
\caption{Plot of $B_2$ versus $x_1$ for the standing (left) and 
traveling (right) circularly polarized Alfv\'en wave problem at
$t=0$ and $t=5$ for the three $(\E^\alpha_z, \E^\circ_z, \E^c_z)$ 
CT algorithms.}
\label{fig:cpaw-1step}
\end{figure}

In T\'oth's analysis \cite{Toth-divB} he found that the errors in the
solution were dominated by errors in the transverse magnetic field
$B_2$ and velocity $v_2$ in our notation.  He presented line plots of
$B_2$ versus $x$ as a function of resolution and numerical scheme.  In
figure \ref{fig:cpaw-1step} we present analogous line plots of $B_2$
versus $x_1$ for the case of a standing and traveling waves including
the initial conditions at $t=0$ and the solutions at $t=5$ for the
three CT algorithms under study.  It is worth noting that in these
plots we have calculated $B_2$ using the cell center magnetic fields
and have included the data for every grid cell in the calculation.
Owing to the angle of the wave with respect to the grid, there are
many data points which fall on a single $x_1$ position.  The lack of
scatter in the data shows that the Alfv\'en wave retains its planar
structure extremely well for the duration of the calculation.

\par
Comparing the plots of $B_2$ versus $x_1$ for the three CT algorithms,
we generally find that the $\E^\circ_z$ or $\E^c_z$ CT algorithm give
nearly identical results, while the $\E^\alpha_z$ CT algorithm is more
dissipative.  In the case of the standing wave solution, the
dissipation rate in the $\E^\alpha_z$ CT algorithm is approximately
twice that of the other two.  In the traveling wave case, the
difference in the dissipation rate is much less indicating that the
dissipation in the first order integration algorithm is comparable.

\par
From these tests, and many additional tests not included here, we
conclude that the $\E^c_z$ CT algorithm has the best dissipation
properties.  The $\E^\alpha_z$ CT algorithm gives stable, yet
diffusive results while the $\E^\circ_z$ CT algorithm appears to have
insufficient dissipation, leading to oscillatory results for advection
problems.  These observations suggest that the $\E^\alpha_z$ CT
algorithm with $\alpha$ set equal to some estimate of the magnitude of
the local gas velocity could also lead to a non-oscillatory CT
algorithm.  This idea is however untested, nor is it clear that it
would result in an algorithm which is superior to the $\E^c_z$ CT
algorithm.  For the remainder of this paper we will use the $\E^c_z$
CT algorithm in test problems.

\section{Second Order CTU + CT Godunov Scheme}
\label{sec:ctu_ct_Godunov}

\par
The corner transport upwind (CTU) method was developed by Colella as
an unsplit, two dimensional sequel to the piecewise parabolic method
(PPM) of Colella and Woodward for Euler's equations.  The algorithm is
second order accurate and degenerates to the base PPM algorithm for
plane-parallel, grid-aligned flows.  Formally, the algorithm can be
described in just a few steps.

\par
First, one calculates left and right interface states at each grid
cell face using the 1-dimensional algorithm from the base PPM scheme.
Using the notation adopted in \S\ref{sec:lr_states}, let these be
denoted by
$(q^{L*}_{i+1/2,j},~q^{R*}_{i+1/2,j},~q^{L*}_{i,j+1/2},~q^{R*}_{i,j+1/2})$.
For MHD, the PPM interpolation used to construct these states must
include all the multidimensional terms identified in
\S\ref{sec:lr_states}.  At each interface one solves the Riemann
problem associated with these interface states and computes the fluxes
$(F^*_{x,i+1/2,j},~F^*_{y,i,j+1/2})$.  Next, one updates the interface
states to the $1/2$ time step
\beq
q^{L}_{i+1/2,j} = q^{L*}_{i+1/2,j} - \frac{1}{2}\frac{\delta t}{\delta y}
\left( F^*_{y,i,j+1/2} - F^*_{y,i,j-1/2} \right)
\label{eq:hydro_lx_state}
\eeq
\beq
q^{R}_{i+1/2,j} = q^{R*}_{i+1/2,j} - \frac{1}{2}\frac{\delta t}{\delta y}
\left( F^*_{y,i+1,j+1/2} - F^*_{y,i+1,j-1/2} \right)
\label{eq:hydro_rx_state}
\eeq
\beq
q^{L}_{i,j+1/2} = q^{L*}_{i,j+1/2} - \frac{1}{2}\frac{\delta t}{\delta x}
\left( F^*_{x,i+1/2,j} - F^*_{x,i-1/2,j} \right)
\label{eq:hydro_ly_state}
\eeq
\beq
q^{R}_{i,j+1/2} = q^{R*}_{i,j+1/2} - \frac{1}{2}\frac{\delta t}{\delta x}
\left( F^*_{x,i+1/2,j+1} - F^*_{x,i-1/2,j+1} \right)
\label{eq:hydro_ry_state}
\eeq
Solving the Riemann problem associated with the four updated interface
states $(q^{L,R}_{i+1/2,j},~q^{L,R}_{i,j+1/2})$ one obtains second
order accurate fluxes which can be used to update $q_{i,j}$ via the
standard finite volume integration relation, equation
(\ref{eq:fv_int}).

\par
Unfortunately, the CTU method as just described is incomplete for MHD
when using constrained transport.  One reason is that equations
(\ref{eq:hydro_lx_state}) - (\ref{eq:hydro_ry_state}) fail to preserve
the ${\bf\nabla\cdot B}=0$ condition.  The marriage of CTU with CT for
MHD requires a modification of equations (\ref{eq:hydro_lx_state}) -
(\ref{eq:hydro_ry_state}) for updating the interface states, and an
additional CT integration step for updating $q_{i,j}^n$ from time
$t^n$ to $t^{n+1}$.

\subsection{Updating the Interface States}
\label{sec:ctu_ct_Interface_Update}

There are two modifications to the CTU method required for MHD using
CT.  The first modification, required by constrained transport, is
that the fluxes $(F^*_{x,i+1/2,j},~F^*_{y,i,j+1/2})$ must be
integrated from face center, to the grid cell corner.  Hence from the
interface centered flux $F^*_{x,i+1/2,j}$ we obtain two corner
centered fluxes $(F^{L*}_{x,i+1/2,j+1/2}, F^{R*}_{x,i+1/2,j-1/2})$
where the superscript $L,R$ indicates that the fluxes have been
integrated from face center to the grid cell corner from either the
left $(-y)$ or right $(+y)$ side of the grid cell corner.  The
labeling of the $y$-fluxes $(F^{L*}_{y,i+1/2,j+1/2},
F^{R*}_{y,i-1/2,j+1/2})$ obtained from $F^*_{y,i,j+1/2}$ follows an
analogous convention.  The second modification, originating from
differences in the form of the equations of ideal MHD when written in
primitive, or conservative variables, is the addition of source terms.
The modified form of the update relations in equations
(\ref{eq:hydro_lx_state}) - (\ref{eq:hydro_ry_state}) for advancing
the interface states to the $1/2$ time step can be formally written as
\beq
q^{L}_{i+1/2,j} = q^{L*}_{i+1/2,j} - \frac{1}{2}\frac{\delta t}{\delta y}
\left( F^{L*}_{y,i+1/2,j+1/2} - F^{L*}_{y,i+1/2,j-1/2} \right)
+ \frac{\delta t}{2} S_{x,i,j}
\label{eq:mhd_lx_state}
\eeq
\beq
q^{R}_{i+1/2,j} = q^{R*}_{i+1/2,j} - \frac{1}{2}\frac{\delta t}{\delta y}
\left( F^{R*}_{y,i+1/2,j+1/2} - F^{R*}_{y,i+1/2,j-1/2} \right)
+ \frac{\delta t}{2} S_{x,i+1,j}
\label{eq:mhd_rx_state}
\eeq
\beq
q^{L}_{i,j+1/2} = q^{L*}_{i,j+1/2} - \frac{1}{2}\frac{\delta t}{\delta x}
\left( F^{L*}_{x,i+1/2,j+1/2} - F^{L*}_{x,i-1/2,j+1/2} \right)
+ \frac{\delta t}{2} S_{y,i,j}
\label{eq:mhd_ly_state}
\eeq
\beq
q^{R}_{i,j+1/2} = q^{R*}_{i,j+1/2} - \frac{1}{2}\frac{\delta t}{\delta x}
\left( F^{R*}_{x,i+1/2,j+1/2} - F^{R*}_{x,i-1/2,j+1/2} \right)
+ \frac{\delta t}{2} S_{y,i,j+1} ~.
\label{eq:mhd_ry_state}
\eeq
The procedure for integrating the fluxes from face center to grid
cell corner, and the need for the source terms will now be described
in turn.

\subsubsection{Integrating the Fluxes to the Grid Cell Corner}
\label{sec:ctu_ct_flux_integrate}

\par
Recall that $f_x(B_y)=-\E_z$ and $f_y(B_x)=\E_z$ and let
$\hat{e}_{B_x}$ and $\hat{e}_{B_y}$ denote the unit vectors for the
$B_x$ and $B_y$ components of the flux vector.  Furthermore, let
$\E^*_{z,i,j+1/2} = \hat{e}_{B_x} \cdot F^{*}_{y,i,j+1/2}$ and
$\E^*_{z,i+1/2,j} = -\hat{e}_{B_y} \cdot F^{*}_{x,i+1/2,j}$.  This
allows us to define
\beq
F^{L*}_{y,i+1/2,j+1/2} = F^{*}_{y,i,j+1/2} + 
(\E^*_{z,i+1/2,j+1/2} - \E^*_{z,i,j+1/2}) \hat{e}_{B_x}
\eeq
\beq
F^{R*}_{y,i+1/2,j+1/2} = F^{*}_{y,i+1,j+1/2} + 
(\E^*_{z,i+1/2,j+1/2} - \E^*_{z,i+1,j+1/2}) \hat{e}_{B_x}
\eeq
\beq
F^{L*}_{x,i+1/2,j+1/2} = F^{*}_{x,i+1/2,j} - 
(\E^*_{z,i+1/2,j+1/2} - \E^*_{z,i+1/2,j}) \hat{e}_{B_y}
\eeq
\beq
F^{R*}_{x,i+1/2,j+1/2} = F^{*}_{x,i+1/2,j+1} - 
(\E^*_{z,i+1/2,j+1/2} - \E^*_{z,i+1/2,j+1}) \hat{e}_{B_y} ~.
\eeq
From a practical point of view, integrating the fluxes from face
center to grid cell corner in this fashion can be thought of as simply
stating that the normal components of the magnetic field at grid cell
interfaces is advanced to time $t^{n+1/2}$ via a CT integral.  Namely, 
\beq
B_{x,i\pm1/2,j}^{n+1/2} = B_{x,i\pm1/2,j}^n 
+ \frac{1}{2} \frac{\delta t}{\delta y}
 \left( \E^*_{z,i\pm1/2,j-1/2} - \E^*_{z,i\pm1/2,j+1/2} \right) 
\eeq
\beq
B_{y,i,j\pm1/2}^{n+1/2} = B_{y,i,j\pm1/2}^n 
- \frac{1}{2} \frac{\delta t}{\delta x}
 \left( \E^*_{z,i-1/2,j\pm1/2} - \E^*_{z,i+1/2,j\pm1/2} \right) ~.
\eeq

\par
The last part of this integration procedure which requires
description is the CT algorithm used to calculate the corner centered
emf $\E^*_{z,i+1/2,j+1/2}$.  To accomplish this, note that the fluxes
$(F^*_{x,i+1/2,j},~F^*_{y,i,j+1/2})$ are equivalent to the fluxes used
in the single step integration algorithm tested in
{}\S\ref{sec:ss_ct_Godunov}.  As such, the calculation of the corner
centered emf $\E^*_{z,i+1/2,j+1/2}$ may be accomplished with any of
the $(\E^\alpha_z, \E^\circ_z, \E^c_z)$ CT algorithms described in
\S\ref{sec:CT_Algorithms}.  For the tests problems
presented in the following sections we will use the $\E^c_z$ CT
algorithm.  We note that following this procedure, the magnetic
fields satisfy the ${\bf\nabla\cdot B}=0$ condition at time
$t^{n+1/2}$.

\subsubsection{Interface State MHD Source Terms}
\label{sec:ctu_ct_source_terms}

\par
The source terms present in equations (\ref{eq:mhd_lx_state}) -
(\ref{eq:mhd_ry_state}) follow from the recognition that if they are
set to zero, the updated interface states are not formally advanced by
$\delta t/2$ for MHD.  The basic reason for this discrepancy lies in
the fact that the interface states as described in
{}\S\ref{sec:lr_states} (and typically implemented in PPM) are
calculated in primitive variables.  As a result,
\beq
q^{L*}_{i+1/2,j} \neq q(x_i + \delta x/2, y_j) 
- \frac{\delta t}{2}\frac{\partial f_x}{\partial x} 
\eeq
and similarly for the other interface states.  To correct this
situation, we define
\beq
S_{x,i,j} \equiv \left (
\begin{array}{c}
0 \\
B^n_{x,i,j} \\
B^n_{y,i,j} \\
B^n_{z,i,j} \\
0 \\
0 \\
v^n_{z,i,j} \\
B^n_{z,i,j} v^n_{z,i,j} 
\end{array} \right) 
\left( \frac{B_{x,i+1/2,j}^n - B_{x,i-1/2,j}^n}{\delta x} \right)
\eeq
and 
\beq
S_{y,i,j} \equiv \left (
\begin{array}{c}
0 \\
B^n_{x,i,j} \\
B^n_{y,i,j} \\
B^n_{z,i,j} \\
0 \\
0 \\
v^n_{z,i,j} \\
B^n_{z,i,j} v^n_{z,i,j} 
\end{array} \right) 
\left( \frac{B_{y,i,j+1/2}^n - B_{y,i,j-1/2}^n}{\delta y} \right) ~.
\eeq
With this choice, the interface states as updated by equations
(\ref{eq:mhd_lx_state}) - (\ref{eq:mhd_ry_state}) include all of the
necessary terms so as to be advanced to time $t^{n+1/2}$.

\par
Note that the choice to include the term $v_z (\partial B_x /
\partial x)$ for $B_z$ (and the associated energy source term) at the
$x$-interfaces in this step, rather than including it when calculating
the interface states as described in {}\S\ref{sec:lr_states} has a
very important consequence; it prevents an erroneous field growth of
$B_z$ in certain circumstances.  To elucidate this situation, consider
a magnetic field loop in the $(x,y)$-plane advected with a uniform
${\bf v} = v_z {\bf \hat{k}}$ and set $B_z=0$ initially.  With the
exception of extrema, $\partial B_x/\partial x$ and $\partial
B_y/\partial y$ are non-zero throughout the field loop. However, since
$v_z$ is uniform the magnetic field $B_z$ should remain equal to zero.
If the term $v_z \partial B_x / \partial x$ is included when
calculating the $x$-interface states in {}\S\ref{sec:lr_states} they
would contain a non-zero $B_z$.  Owing to the coherent structure of
the in-plane field loop, the the values of $B_z$ in the interface
states will also have a coherent structure.  Upon updating the
interface states with the transverse flux gradients, the growth of
$B_z$ is diminished, however it is not canceled identically.  The net
result is an unphysical growth of a coherent $B_z$ which eventually
influences the in-plane dynamics.  The choice of source terms
described in section {}\S\ref{sec:lr_states} and this section
maintains $B_z$ to the level of roundoff error with an incoherent
structure.  Hence, the algorithm presented here accurately captures
the balance of the terms proportional to $\partial B_x / \partial x$
and $\partial B_y / \partial y$, as described in
{}\S\ref{sec:lr_states}, in both the predictor and corrector steps for
calculating the interface state values of $B_z$.

\subsection{The Constrained Transport Update Algorithm}
\label{sec:ctu_ct_update_alg}

After having updated the interface states to time $t^{n+1/2}$ via
equations (\ref{eq:mhd_lx_state}) - (\ref{eq:mhd_ry_state}), the
interface flux calculation is repeated giving rise to the second order
accurate fluxes $(F_{x,i+1/2,j}^{n+1/2},~F_{y,i,j+1/2}^{n+1/2})$.  In the CTU
algorithm, this set of fluxes is used to evolve $q^n_{i,j}$ to time
$t^{n+1}$. However, in order to evolve the magnetic fields via
constrained transport, we must extend the CT algorithms described in
{}\S\ref{sec:CT_Algorithms}.  Requiring that the algorithm reduce to
the base integration algorithm for plane-parallel, grid-aligned flows
we find that we simply need to advance the electric field gradient
calculation to the half time step, i.e. equation
(\ref{eq:Ez_gradient}) is replaced with
\beq
\left(\frac{\partial \E_z}{\partial x} \right)_{i+1/4,j}^{n+1/2} = 
\frac{2}{\delta x} \left( \E_{z,i+1/2,j}^{n+1/2} - \E_{z,i,j}^{n+1/2} \right) ~.
\label{eq:Ez_gradient_hdt}
\eeq
The electric field $\E_{z,i,j}^{n+1/2}$ is the cell center value
advanced by $\delta t/2$, that is
\beq
\E_{z,i,j}^{n+1/2} = v_{y,i,j}^{n+1/2}B_{x,i,j}^{n+1/2}
- v_{x,i,j}^{n+1/2}B_{y,i,j}^{n+1/2}
\label{eq:cell-centered-E}
\eeq
where to be consistent with the integration
scheme, the cell center magnetic fields are given by
\begin{eqnarray}
B_{x,i,j}^{n+1/2} & = &
\frac{1}{2} \left( B_{x,i-1/2,j}^{n+1/2} + B_{x,i+1/2,j}^{n+1/2} \right) 
\label{eq:Bx-hs} \\
B_{y,i,j}^{n+1/2} & = &
\frac{1}{2} \left( B_{y,i,j-1/2}^{n+1/2} + B_{y,i,j+1/2}^{n+1/2} \right)
\label{eq:By-hs}
\end{eqnarray}
where the field components on the right hand side of these equations
are equal to the normal components of the magnetic field in the
interface states $(q^{L,R}_{i\pm1/2,j},~q^{L,R}_{i,j\pm1/2})$.  The
density, $x$- and $y$-momenta needed to compute the velocity
components in equation (\ref{eq:cell-centered-E}) are advanced by
$\delta t/2$ using
\beq
q_{i,j}^{n+1/2} = q_{i,j}^n + \frac{1}{2} \frac{\delta t}{\delta x}
\left( F_{x,i-1/2,j}^* - F_{x,i+1/2,j}^* \right) + \frac{1}{2}
\frac{\delta t}{\delta y} \left( F_{y,i,j-1/2}^* - F_{y,i,j+1/2}^*
\right)
\label{eq:q-hs}
\eeq
where the fluxes are those calculated in the first step of the CTU
algorithm.

\subsection{Summary}

\par
The following steps summarize the CTU + CT algorithm for MHD:
\begin{enumerate}
\item Calculate the $x$- and $y$-interface states 
$(q^{L*}_{i+1/2,j},~q^{R*}_{i+1/2,j},~q^{L*}_{i,j+1/2},~q^{R*}_{i,j+1/2})$
using the PPM algorithm, and the multidimensional source terms as
described by equations (\ref{eq:wave-prop-src}) -
(\ref{eq:wave-prop-matrix}) in {}\S\ref{sec:lr_states}.
\item Calculate the $x$- and $y$-interface fluxes 
$(F^*_{x,i+1/2,j},~F^*_{y,i,j+1/2})$ associated with the interface states
$(q^{L*,R*}_{i+1/2,j},~q^{L*,R*}_{i,j+1/2})$
via a Riemann solver.
\item Using the $\E^c_z$ CT algorithm described in equations 
(\ref{eq:2d_CT_Ez}) and (\ref{eq:Ndup_CT_Ez}) integrate the face
centered fluxes to the grid cell corner as described in
\S\ref{sec:ctu_ct_flux_integrate}.
\item Compute the the four updated interface states 
$(q^{L,R}_{i+1/2,j},~q^{L,R}_{i,j+1/2})$ via equations
(\ref{eq:mhd_lx_state}) - (\ref{eq:mhd_ry_state}) with the source terms
detailed in {}\S\ref{sec:ctu_ct_source_terms}.
\item Compute the $x$- and $y$-interface fluxes 
$(F^{n+1/2}_{x,i+1/2,j},~F^{n+1/2}_{y,i,j+1/2})$ associated with the
interface states $(q^{L,R}_{i+1/2,j},~q^{L,R}_{i,j+1/2})$ via a
Riemann solver.
\item Compute the grid cell corner centered electric field 
$\E_{z,i+1/2,j+1/2}^{n+1/2}$ using the $\E^c_z$ CT algorithm described
in equations (\ref{eq:2d_CT_Ez}) and (\ref{eq:Ndup_CT_Ez}) advanced to
time $t^{n+1/2}$ as described in \S\ref{sec:ctu_ct_update_alg}.
\item Advance the surface averaged normal components of the magnetic 
field from time $t^n$ to $t^{n+1}$ using equations (\ref{eq:ct_Bx_int})
and (\ref{eq:ct_By_int}).
\item Advance the remaining volume averaged conserved quantities from 
time $t^n$ to $t^{n+1}$ using equation (\ref{eq:fv_int}).
\end{enumerate}
This completes the description of the algorithm.  It is second order
accurate, unsplit, and preserves the ${\bf\nabla\cdot B} = 0$
constraint throughout the time step.  In the following section we
apply this CTU + CT scheme to a variety of test problems.

\section{Tests}
\label{sec:tests}

In this section we present results obtained with the CTU + CT
integration algorithm just described.  Throughout these tests we use
the $\E_z^c$ CT algorithm.

\subsection{Field Loop Advection}

The advection of a magnetic field loop discussed in
{}\S\ref{sec:ss_ct_Godunov:field_loop} was instructive for assessing
the dissipation in the different CT algorithms.  In this subsection we
present the results obtained for this problem with the second-order
CTU + CT algorithm.  The grid resolution, and initial conditions are
equivalent to those used in {}\S\ref{sec:ss_ct_Godunov:field_loop}.

\par
\begin{figure}[hbt]
\begin{center}
\includegraphics*[width=2.5 in]{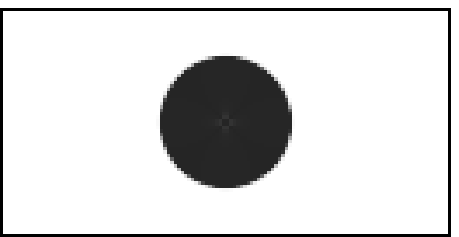} \hfill 
\includegraphics*[width=2.5 in]{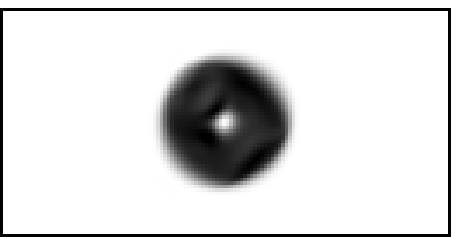} 
\end{center}
\caption{Grey-scale images of the magnetic pressure $(B_x^2 + B_y^2)$
at $t=0$ (left) and $t=2$ (right) using the CTU + CT integration
algorithm.}
\label{fig:Loop-B2-ctu}
\end{figure}
\begin{figure}[hbt]
\begin{center}
\includegraphics*[width=4.5 in]{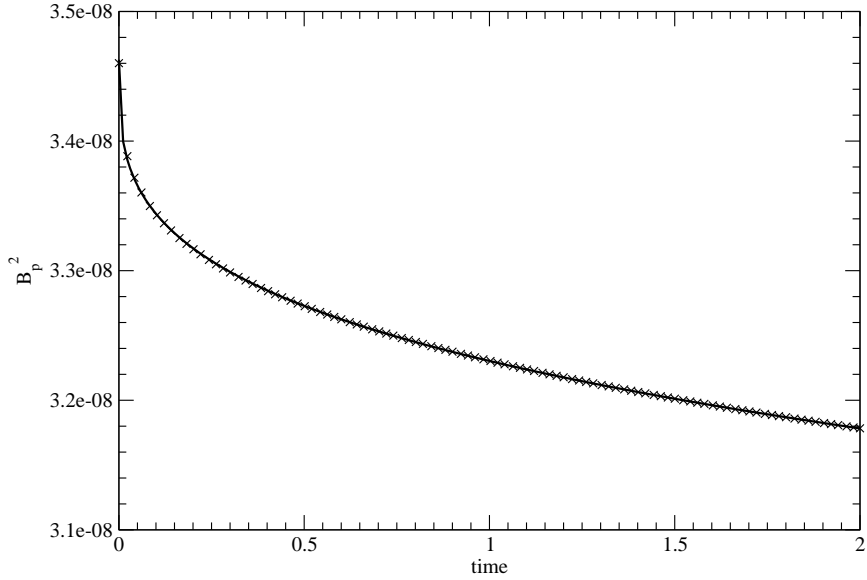}
\end{center}
\caption{Plot of the volume averaged magnetic energy density $B^2$ as 
a function of time.  The solid line is a power law curve fit to the
data points denoted by symbols.}
\label{fig:Loop-B2-Decay}
\end{figure}

In figure \ref{fig:Loop-B2-ctu} we present grey-scale images of $B^2$
at times $t=0$ and $t=2$.  Comparing these figures we find that the
majority of the field dissipation has occurred at the center and
boundaries of the field loop, where the current density is initially
singular.  A more quantitative measure of the magnetic field
dissipation rate is given by the time evolution of the volume average
of $B^2$ as shown in figure {}\ref{fig:Loop-B2-Decay}.  We find that
the measured values (denoted by symbols) is well described by a power
law (solid line) of the form $B^2 = A \left(1-(t/\tau)^\alpha \right)$
with $A=3.463\times10^{-8}$, $\tau=10.614\times 10^3$ and
$\alpha=0.2914$.

\par
\begin{figure}[hbt]
\begin{center}
\includegraphics*[width=2.5 in]{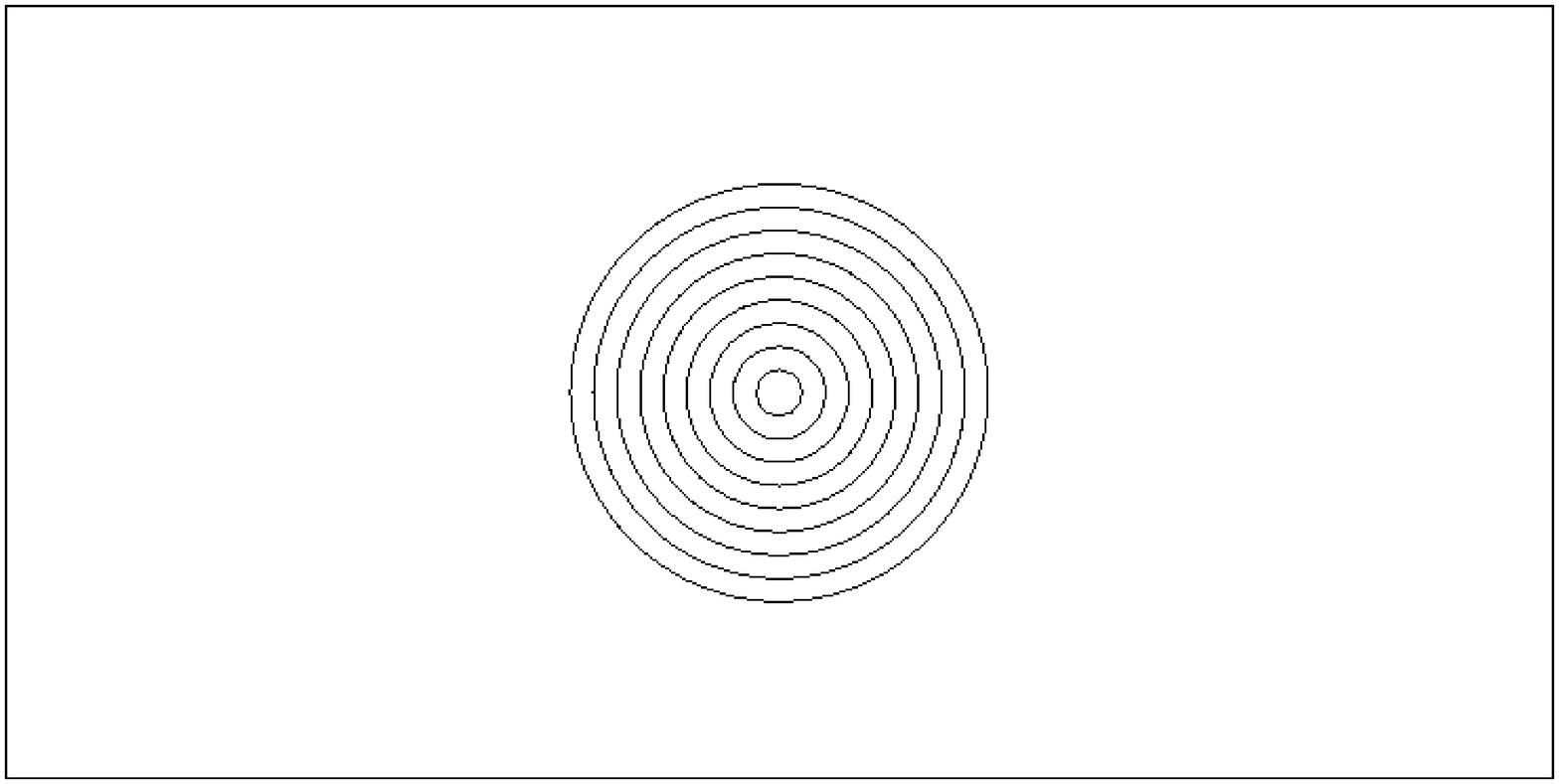} \hfill
\includegraphics*[width=2.5 in]{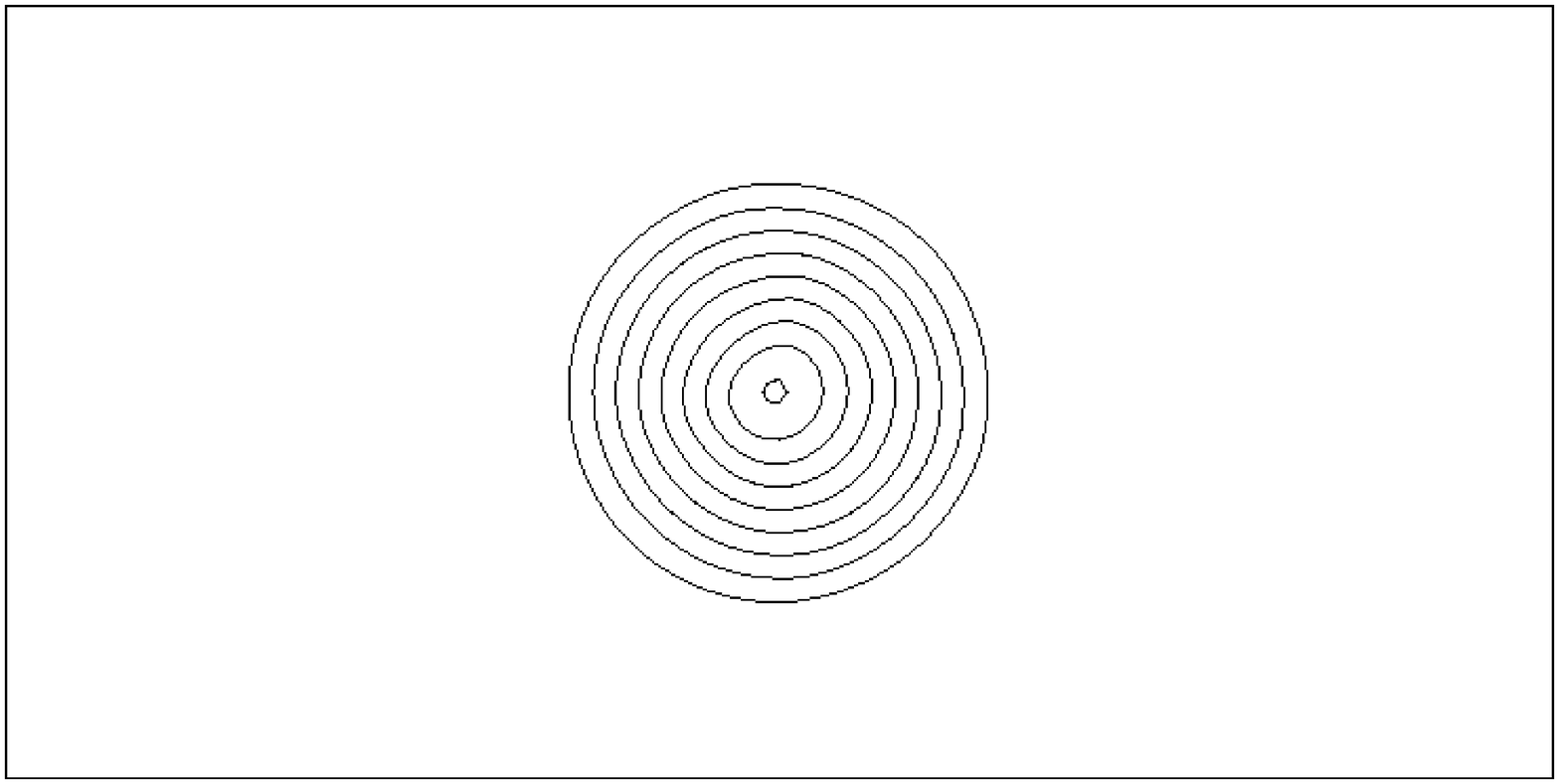} 
\end{center}
\caption{Magnetic field lines at $t=0$ (left) and $t=2$ (right) 
using the CTU + CT integration algorithm.}
\label{fig:Loop-B-lines-ctu}
\end{figure}

Another important indicator of the properties of the integration
algorithm is the geometry of the magnetic field lines.  Note that
since the CT method evolves the interface magnetic flux (preserving
${\bf \nabla \cdot B}=0$) one may readily integrate to find the
$z$-component of the magnetic vector potential.  The magnetic field
lines presented in figure {}\ref{fig:Loop-B-lines-ctu} are obtained by
contouring $A_z$.  The same values of $A_z$ are used for the contours
in both the $t=0$ and the $t=2$ images.  By $t=2$ the inner most field
line has dissipated.  It is quite pleasing, however, to note that the
CTU + CT algorithm preserves the circular shape of the magnetic field
lines, even at this low resolution.

%

\subsection{Circularly Polarized Alfv\'en Wave}

The test problem involving the propagation of circularly polarized
Alfv\'en waves at an oblique angle to the grid was described in
{}\S\ref{sec:ss_ct_Godunov:cp_alfven}.  In this subsection we present
a resolution study for both standing and traveling Alfv\'en waves.
The initial conditions are equivalent to those used in
{}\S\ref{sec:ss_ct_Godunov:cp_alfven} only with $N=\{4,~8,~16,~32\}$.

\par
\begin{figure}[hbt]
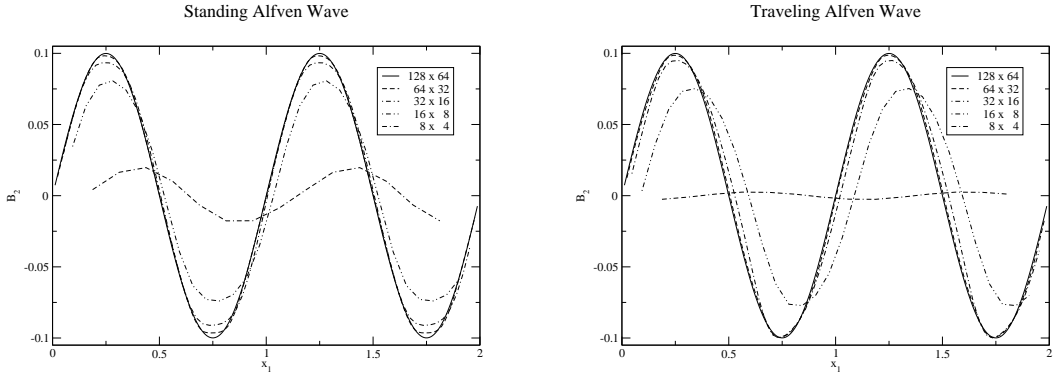

\begin{center}
\includegraphics*[width=2.5in]{cp_alfven.standing.eps} \hfill
\includegraphics*[width=2.5in]{cp_alfven.traveling.eps}
\end{center}
\caption{Plot of $B_2$ versus $x_1$ at $t=5$ for the standing (left) and 
traveling (right) circularly polarized Alfv\'en wave problem.  For
comparison, the initial conditions at $t=0$ for the $N=64$ case is
also included.}
\label{cpaw-test}
\end{figure}

As a diagnostic of the solution accuracy, we plot the in-plane
component of the magnetic field, $B_2$, perpendicular to the wave
propagation direction, $x_1$, in figure \ref{cpaw-test}.  These plots
are constructed using the cell center components of the magnetic
field, and each grid cell is included in the plots.  Hence, the lack
of scatter demonstrates that the solutions retain their planar
symmetry quite well.  Figure \ref{cpaw-test} includes the solutions at
time $t=5$ with $N=\{4,~8,~16,~32\}$ for both standing and traveling
waves.  For comparison, we also include the initial conditions for the
$N=64$ case.  We find that these results compare well against other
published calculations {}\cite{Toth-divB,Pen-MHD-03} and we find no
indication of the parametric instability for the parameters adopted
here.
%

\subsection{Rotated Shock Tube Problem}

\par
The solution to the one dimensional Riemann problem has long been used
as a test of numerical algorithms \cite{Sod,Liska-Wendroff-Test}.
Solving the same problem in a two dimensional domain with the
initially planar discontinuity rotated by some angle with respect the
the grid can also be a robust test of the integration algorithm.  In
addition to the usual questions, one is also interested in how well
the planar symmetry is preserved for flows which are oblique to the
grid.

\par
For MHD this implies a particularly stringent condition on the
component of the magnetic field in the direction of the initial
discontinuity normal.  Using the coordinate transformations in
equations (\ref{eq:coord_trans_x}) - (\ref{eq:coord_trans_z}), let the
initial discontinuity lie in the plane $x1=$ constant.  Then for MHD,
the solution to the one-dimensional Riemann problem should have $B_1=$
constant, which requires a balance between the $x$- and $y$-gradients
of $\E_z$ such that $\partial \E_z / \partial x_2 = 0$.  T\'oth
\cite{Toth-divB} has recently shown that in some cases, schemes which
do not preserve the ${\bf\nabla\cdot B}=0$ condition can result in a
solution in which $B_1$ contains a jump across a shock; see also
\cite{Falle98}.

\par
In the trivial case, rotated shock tube problems are initialized with
the shock tube discontinuity oriented at a 45 degree angle with
respect to the grid, i.e. with a coordinate rotation angle
$\theta=\tan^{-1}(\delta x/\delta y)$.  Examples of test calculations
performed using this configuration can be found in
{}\cite{Dai-Woodward-98,RMJF98,Falle98}.  We have run a variety of
shock tube problems with this configuration (with both $\delta x =
\delta y$ and $\delta x \neq \delta y$) and find that in all of our
tests, the parallel component of the magnetic field, $B_1$, remains
equal to a constant with variations which are of the order of roundoff
error.  We assert that this is a result of the symmetry of the initial
conditions with respect to the grid.

\par
A non-trivial configuration with a coordinate rotation angle of
$\theta=\tan^{-1}(2) \approx 63.4^\circ$ and $\delta x = \delta y$ was
recently suggested by T\'oth \cite{Toth-divB} and has been adopted
elsewhere {}\cite{Londrillo-DelZanna04,Crockett} as well.  This
problem is more challenging because at the discrete, grid-scale level
the initial conditions contain variations along the plane of the
initial shock tube discontinuity.  Moreover, the symmetry in this
configuration is such that $q_{i,j} = q_{i+2,j-1}$ which is outside of
the integration stencil for most integration algorithms.  This is
especially true in the neighborhood of shocks where most integration
algorithms drop to first order.  Typically, it is in the neighborhood
of shocks where one find oscillations, or in some cases even jumps, in
$B_1$.

\par
We choose to simulate rotated shock tube problems on a grid of $N_x
\times N_y$ grid cells with $\delta x = \delta y$ and the shock tube
discontinuity oriented along the grid diagonal.  Let $C$ equal the
greatest common divisor of $(N_x,~N_y)$ and define $r_x \equiv N_x/C$
and $r_y \equiv N_y/C$.  With this configuration, the coordinate
rotation angle $\theta=\tan^{-1}(r_x/r_y)$ and the symmetry is such
that $q_{i,j} = q_{i+r_x,j-r_y}$.  Note that this computational grid
can also be described as containing $C \times C$ ``macro-cells'' each
of which is $r_x \times r_y$ grid cells in size.  We have run a
variety of shock tube problems with $(r_x,~r_y)~=$ $(2,~1)$, $(3,~2)$,
$(5,~4)$, etc. and in all cases find results which are mutually
consistent.

\par
In the interest of presenting solutions which can be compared to
previously published results
{}\cite{Londrillo-DelZanna04,Crockett,Toth-divB} we will now focus on
the $(r_x,~r_y)~=$ $(2,~1)$ case with $N_x=256$ and $N_y=128$.  The
particular problem studied has a left state given by
$V^L=(1,~10,~0,~0,~5/\sqrt{4\pi},~5/\sqrt{4\pi},0,20)$ and a right
state given by $V^R=(1,~-10,~0,~0,~5/\sqrt{4\pi},~5/\sqrt{4\pi},0,1)$
where $V=(\rho,~v_1,~v_2,~v_3,~B_1,~B_2,~B_3,~P)$.  Among other
places, the one dimensional solution to this Riemann problem can be
found in figure 1a of {}\cite{Ryu-Jones-95}.

\par
\begin{figure}[htb]
\begin{center}
\includegraphics*[width=4in]{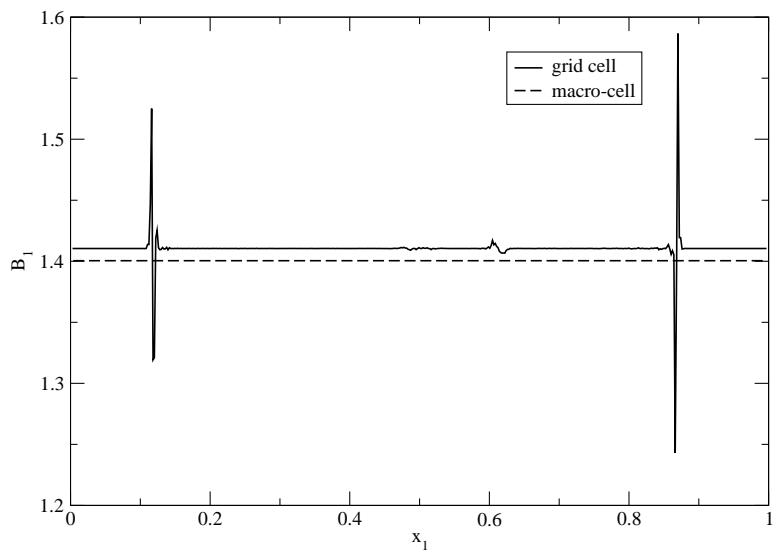}
\end{center}
\caption{Plot of $B_1$ versus $x_1$ for the $(r_x,~r_y)~=$ $(2,~1)$ case at
time $t=0.08$ using the grid cell centered ${\bf B}$ and the macro-cell
centered ${\bf B}$.  The data for the macro-cell centered ${\bf B}$ has been
offset vertically by $0.01$ to for clarity.}
\label{fig:RJ1a-line-plot}
\end{figure}

In figure \ref{fig:RJ1a-line-plot} we present line plots of the
parallel component of the magnetic field $B_1$ versus the parallel
coordinate $x_1$.  These line plots include every point in the
computational domain, hence the lack of scatter indicates that the
solution retains the planar structure quite well.  The first line
plot, labeled ``grid cell'', is constructed using the cell center
magnetic field components.  We find oscillations in $B_1$ which are
roughly $10\%$ of $B_1$, with the largest oscillations occurring at
the fast-mode shocks and weaker oscillations at the left and right
propagating slow-mode rarefaction and shock respectively.  We note
that the $\E^\alpha_z$ CT algorithm does not reduce these oscillations
further when compared to the $\E^\circ_z$ CT algorithm.  Hence the
oscillations in $B_1$ are not a result of insufficient dissipation in
the CT algorithm.  The second line plot, labeled ``macro-cell'', is
constructed by first conservatively averaging the solution onto a grid
of $128 \times 128$ ``macro-cells'' before computing the macro-cell
center component of $B_1$.  The variations in $B_1$ when averaged onto
a macro-cell are of the order of roundoff error.  Note that we obtain
the same result for other rotation angles, e.g. with $(r_x,~r_y)~=$
$(3,~2)$, $(5,~4)$, etc.  We conclude that the oscillations in $B_1$
versus $x_1$ are a simple consequence of the fact that on the scale of
grid cells, the discretized solution contains variations in the
$x_2$-direction.  Upon averaging the solution onto the macro-cells,
this variation is eliminated, and we recover the condition $B_1=$ a
constant.  Note that this also suggests that if one wishes to
eliminate the oscillations in $B_1$ it would require a viscosity with
a stencil whose size is at least as large as the macro-cell.

\par
The recovery of $B_1=$ a constant upon averaging the solution onto a
grid of macro-cells is clearly consistent with magnetic flux
conservation and plane parallel symmetry, yet it does not appear to be
a trivial result.  For example, it is clear that schemes which
generate a jump in $B_1$ \cite{Toth-divB,Falle98} can not recover this
result.

%

\subsection{Linear Wave Convergence}
\label{sec:linear_wave}

In this subsection we show that the CTU + CT integration algorithm
converges with second order accuracy for linear amplitude waves.  The
computational domain extends from $0\le x \le 2/\sqrt{5}$, and $0 \le y
\le 1/\sqrt{5}$, is resolved on a $2N \times N$ grid and has periodic
boundary conditions on both $x$- and $y$-boundaries. The linear wave
propagates at an angle $\theta = \tan^{-1}(2) \approx 63.4^\circ$ with
respect to the $x$-axis and has a wavelength $\lambda=2/5$.  Using the
coordinate rotation described by equations (\ref{eq:coord_trans_x}) -
(\ref{eq:coord_trans_z}), the initial conserved variable state vector
is given by
\beq
q^0 = \bar{q} + \varepsilon R_k \cos(2 \pi x_1)
\eeq
where $\bar{q}$ is the mean background state, $\varepsilon=10^{-6}$ is
the wave amplitude, and $R_k$ is the right eigenvector in conserved
variables for wave mode $k$ (calculated in the state $\bar{q}$).  In
order to enable others to perform the same tests presented here and
compare the results in a quantitative manner, we include the numerical
values for the right eigenvectors in the appendix. As in previous 2D
calculations, the in-plane components of the magnetic field
$(B_x,B_y)$ are initialized via the $z$-component of the magnetic
vector potential.

\par
The mean background state $\bar{q}$ is selected so that the wave
speeds are well separated and there are no inherent symmetries in the
magnetic field orientation.  It is most convenient to describe it in
terms of the associated primitive variables and in the rotated
coordinate system given by equations (\ref{eq:coord_trans_x}) -
(\ref{eq:coord_trans_z}).  The density $\bar{\rho}=1$ and gas pressure
$\bar{P}=1/\gamma=3/5$.  The velocity component parallel to the wave
propagation direction, $\bar{v}_1=1$ for the entropy mode test and
$\bar{v}_1=0$ for all other wave modes.  The transverse velocity
components $\bar{v}_2=\bar{v}_3=0$.  The magnetic field components
$\bar{B}_1=1$, $\bar{B}_2=\sqrt{2}$, and $\bar{B}_3=1/2$.  With this
choice, the slow mode speed $c_s=1/2$, the Alfv\'en speed $c_a=1$, and
the fast mode speed $c_f=2$ in the wave propagation direction.

\par
\begin{figure}[hbt]
\begin{center}
\includegraphics*[width=4.0in]{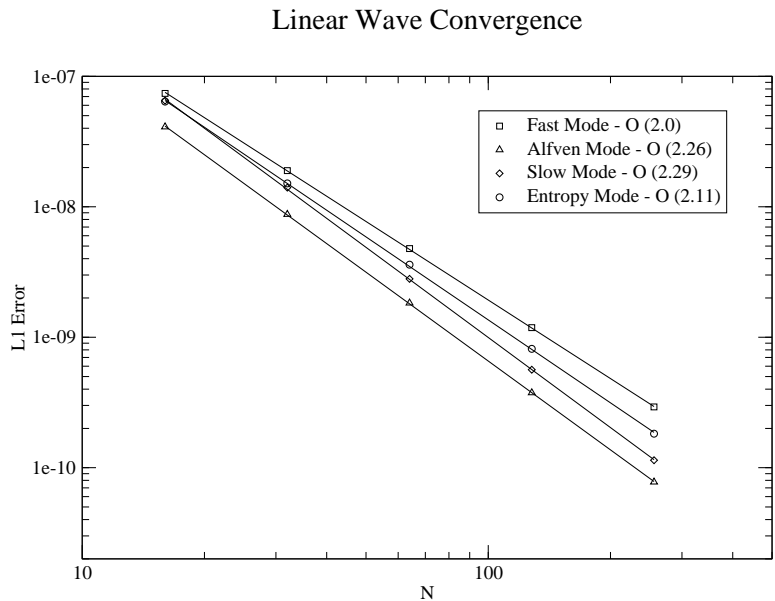}
\end{center}
\caption{Linear wave convergence of fast, Alfv\'en, slow and entropy modes
using the CTU + CT integration algorithm.  The symbols denote the
calculated L1 error norm.  The lines are power law curve fits to the
data and the order of convergence for each wave mode is indicated in
the legend.}
\label{fig:LinWaveError}
\end{figure}

The error in the solution is calculated after propagating the wave for
a distance equal to 1 wavelength.  Hence, the initial state is evolved
for a time $t=\lambda/c$ where $c$ is the speed of the wave mode under
consideration.  For each component $k$ of the conserved variable
vector $q$ we calculate the L1 error with respect to the initial
conditions 
\beq
\delta q_k = \frac{1}{2N^2} \sum_i \sum_j |q_{i,j,k}^n - q_{i,j,k}^0|
\eeq
by summing over all grid cells $(i,j)$.  We use the cell center
components of the in-plane magnetic field components $(B_x,B_y)$ in
computing this error.  In figure \ref{fig:LinWaveError} we plot the
norm of this error vector 
\beq
\| \delta q \| = \sqrt{ \sum_k (\delta q_k)^2 } 
\eeq
for the fast, Alfv\'en, slow and entropy modes.  This plot shows that
the solution for each wave mode converges with at least second order
accuracy.  The order of convergence for each wave mode, obtained by a
power law fit to the errors, is indicated in the legend of figure
{}\ref{fig:LinWaveError}.  We note in passing that if the interface
state reconstruction algorithm is performed using piecewise linear
interpolation, instead of piecewise quadratic, the error is
proportional to $N^{-2}$ for all wave modes and the amplitude is
increased slightly.

%

\subsection{Current Sheet}

In this subsection we present a problem which is particularly
sensitive to the numerical dissipation and demonstrates of the
robustness of the integration algorithm.  The computational domain
extends from $0\le x \le 2$, and $0 \le y \le 2$, is resolved on an $256
\times 256$ grid and has periodic boundary conditions on both $x$- and
$y$-boundaries.  The density $\rho=1$ and the magnetic field
components $B_x = B_z = 0$ and
\beq
B_y =  \left \{
\begin{array}{ll}
B_0  & \textrm{ if $0 \le x < 1/2$} \\
-B_0 & \textrm{ if $1/2 \le x \le 3/2$} \\
B_0  & \textrm{ if $3/2 < x \le 2$}
\end{array}
\right. 
\eeq
where $B_0=1$.  Hence there are initially two current sheets in the
computational domain and the characteristic Alfv\'en speed $c_a =
B_0/\sqrt{\rho}=1$.  The gas pressure $P=0.1$ such that $\beta = 2
P/B_0^2 = 0.2$ and the dynamics are initially magnetically dominated.
The ratio of the Alfv\'en
speed to the sound speed $B_0/\sqrt{\gamma P} \approx 2.45$, hence
magnetically driven dynamics are supersonic.  The initial velocity
components $v_x = v_0\sin(\pi y)$ with $v_0=0.1$, $v_y = v_z = 0$.
For $v_0/c_a \ll 1$ the ensuing dynamics are well characterized by
linear Alfv\'en waves.  For the values selected here this is
approximately true at early times, until magnetic reconnection and
nonlinear effects come to influence the dynamics.

\par
\begin{figure}[hbt]
\begin{center}
\includegraphics*[width=1.5in]{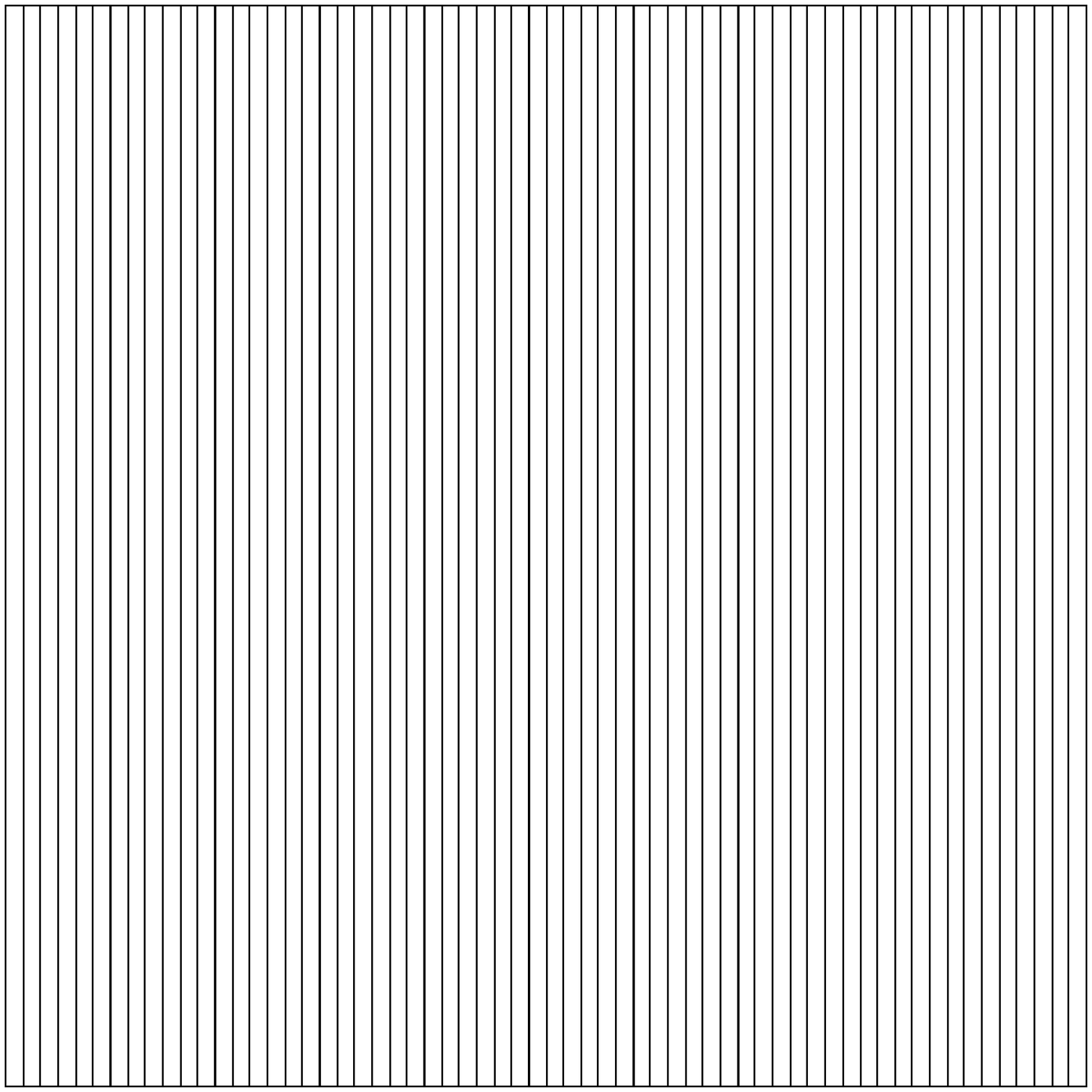} \hfill
\includegraphics*[width=1.5in]{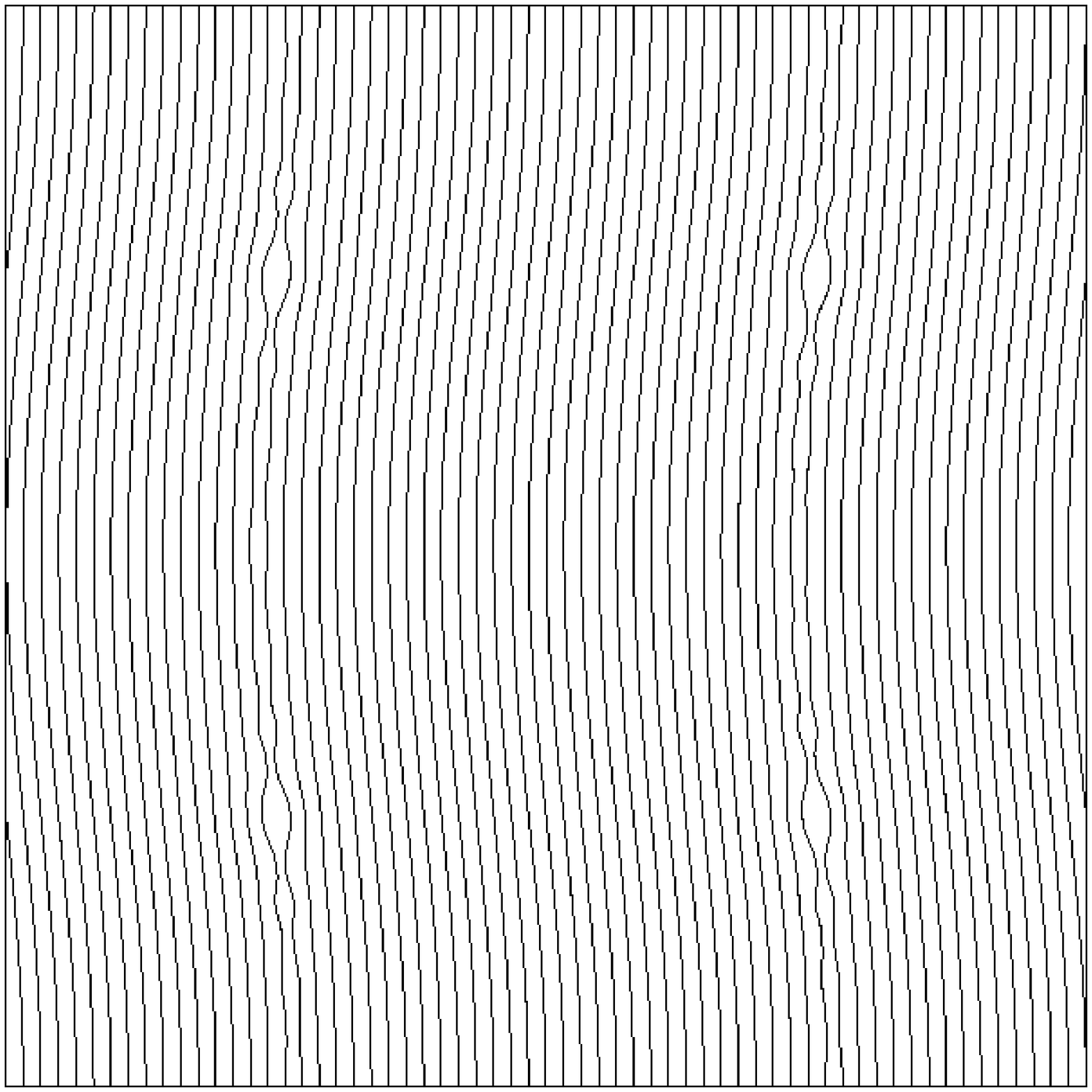} \hfill
\includegraphics*[width=1.5in]{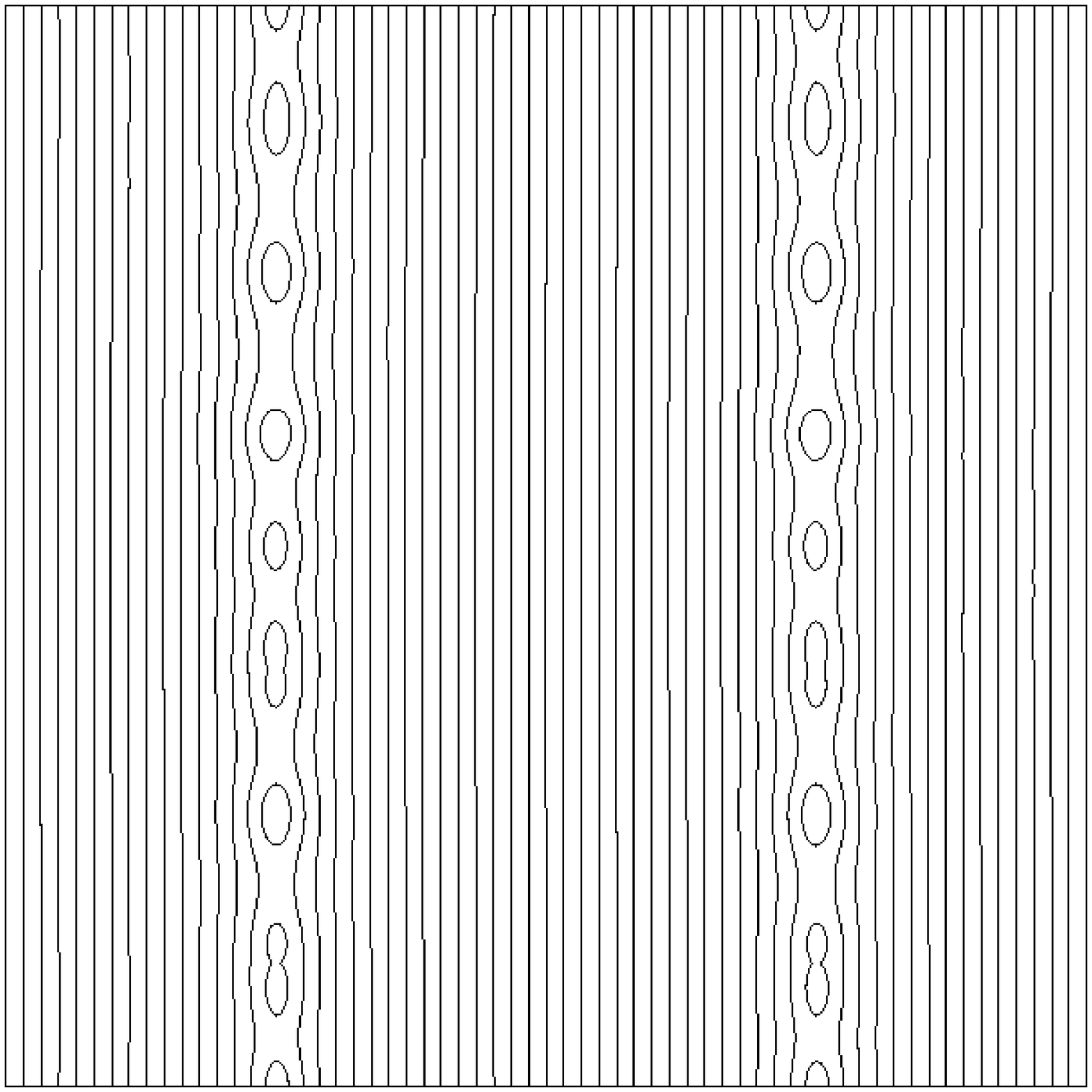} \\[0.25in]
\includegraphics*[width=1.5in]{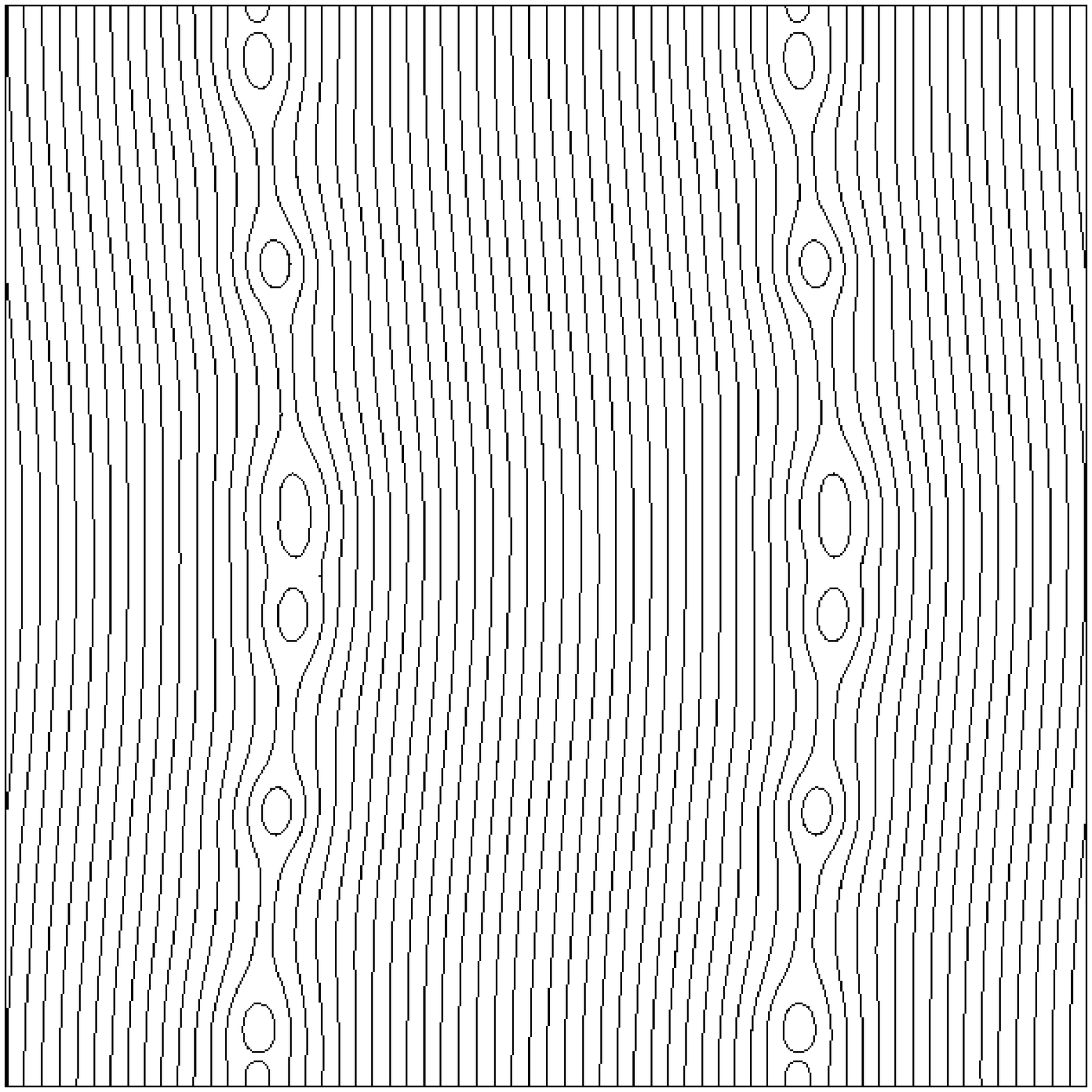} \hfill
\includegraphics*[width=1.5in]{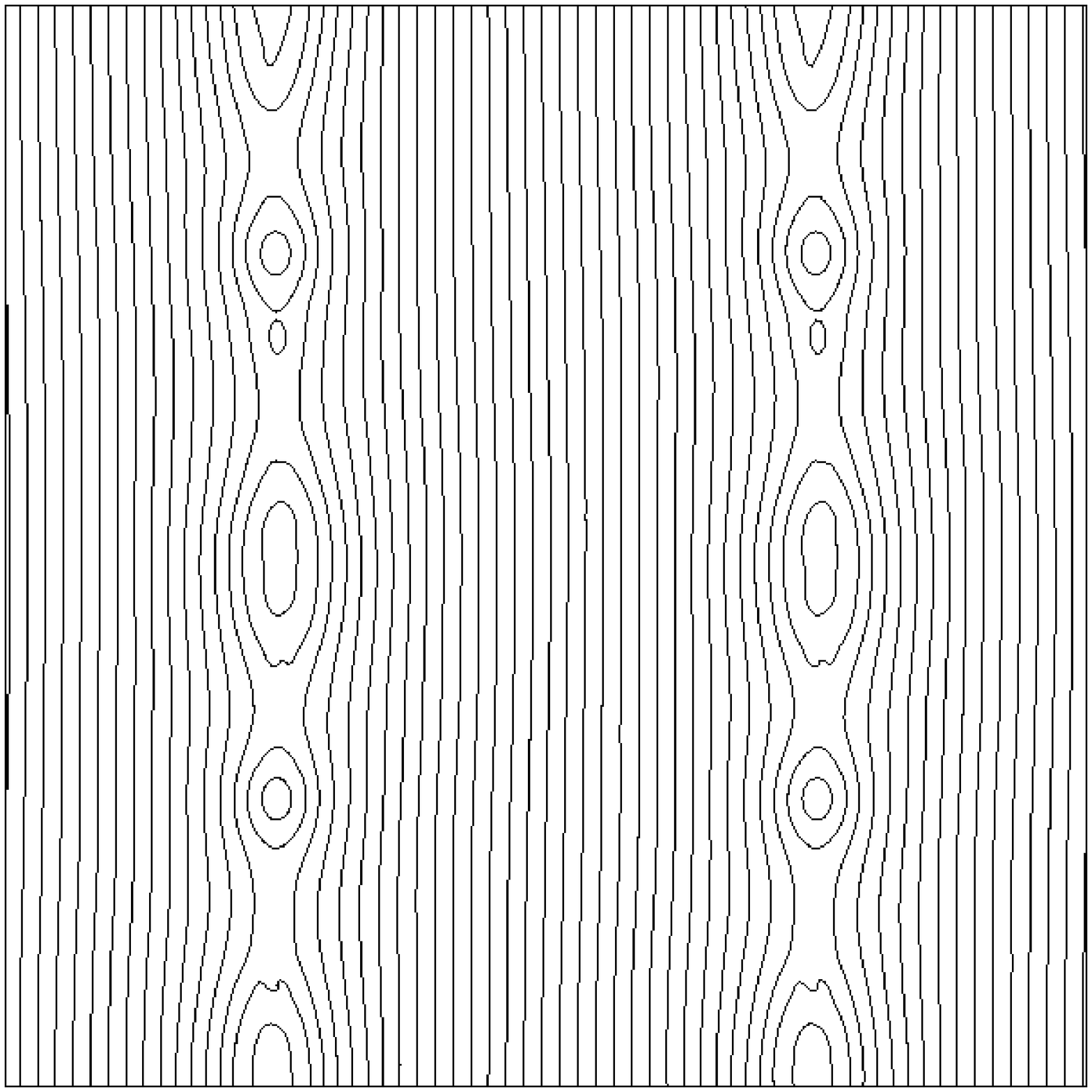} \hfill
\includegraphics*[width=1.5in]{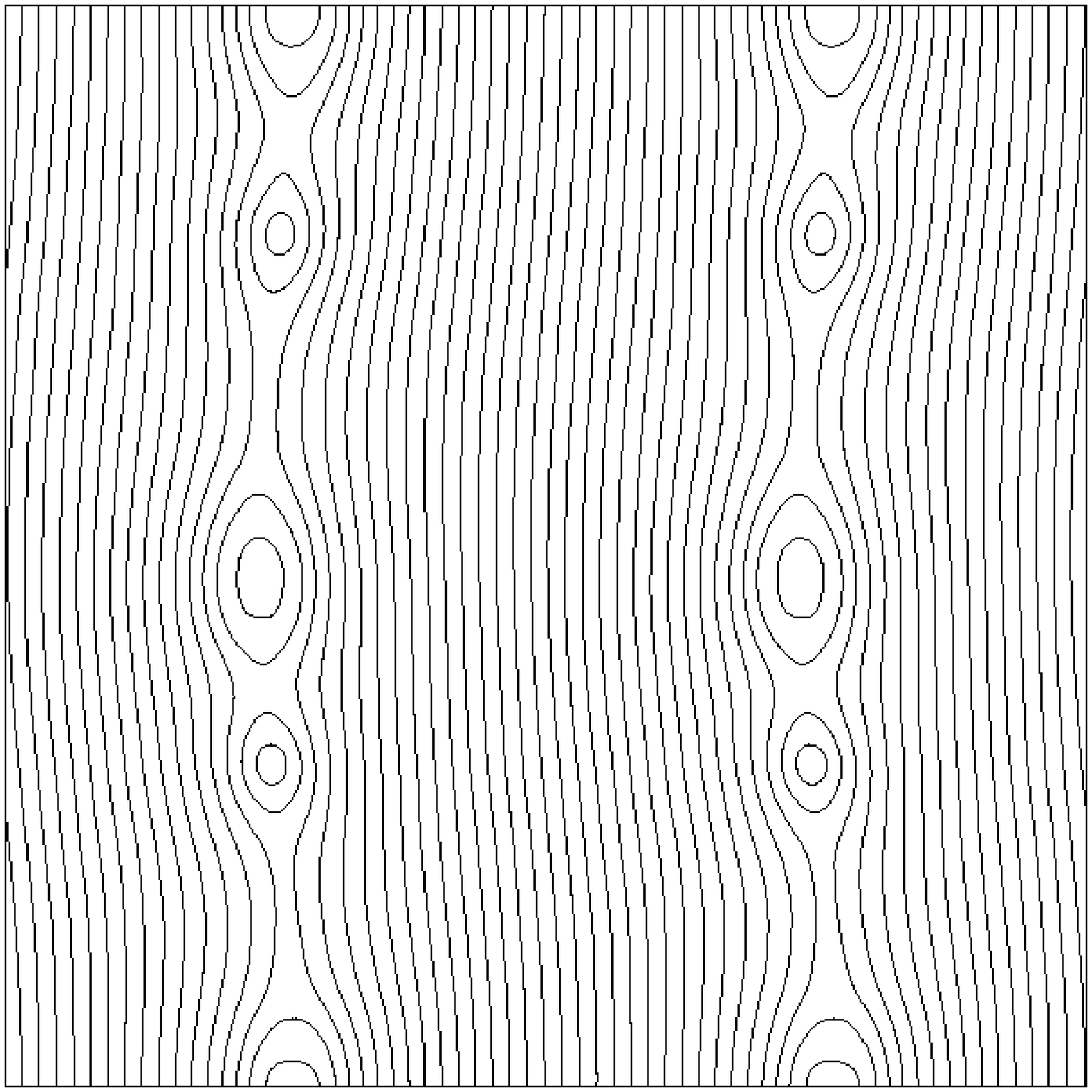} \\[0.25in]
\includegraphics*[width=1.5in]{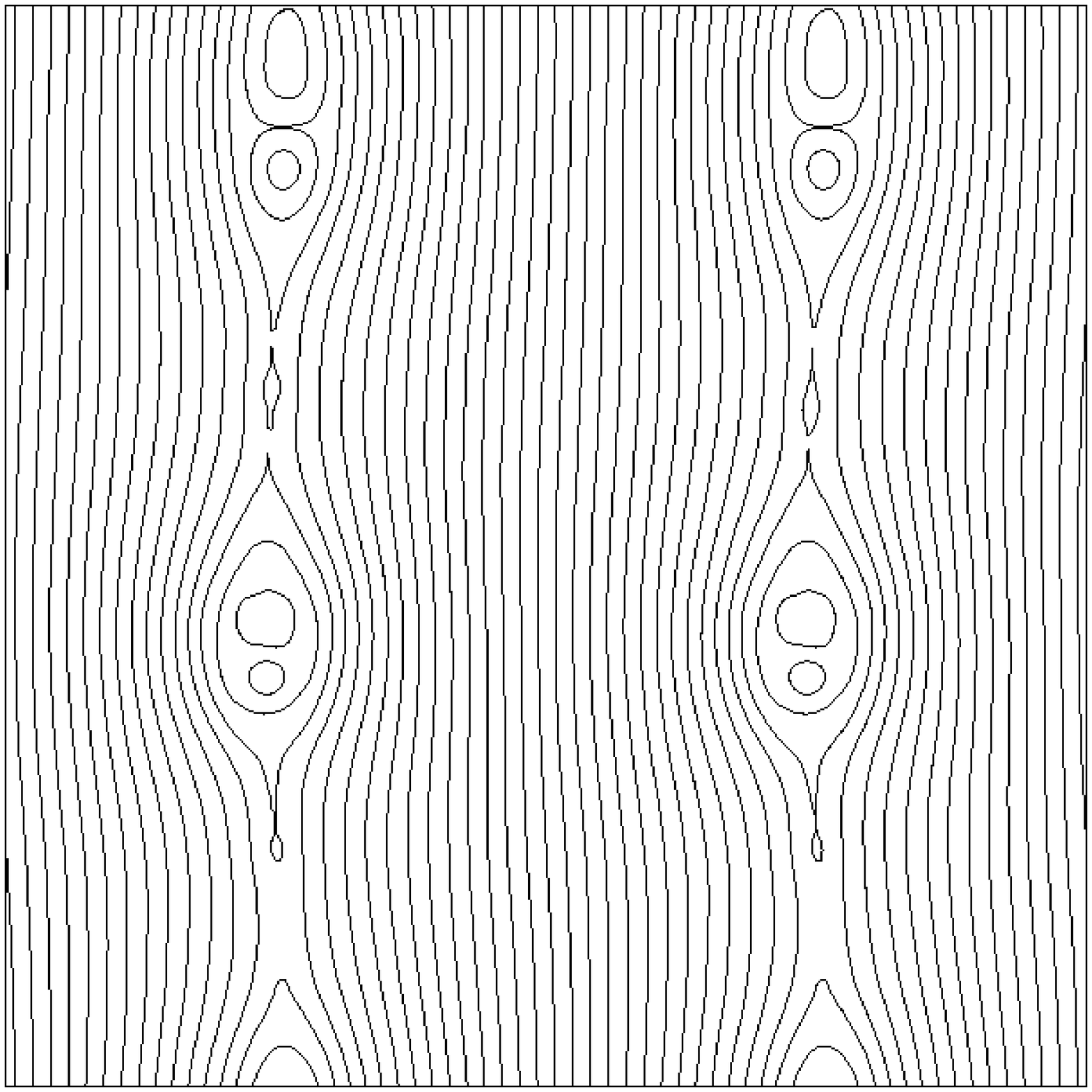} \hfill
\includegraphics*[width=1.5in]{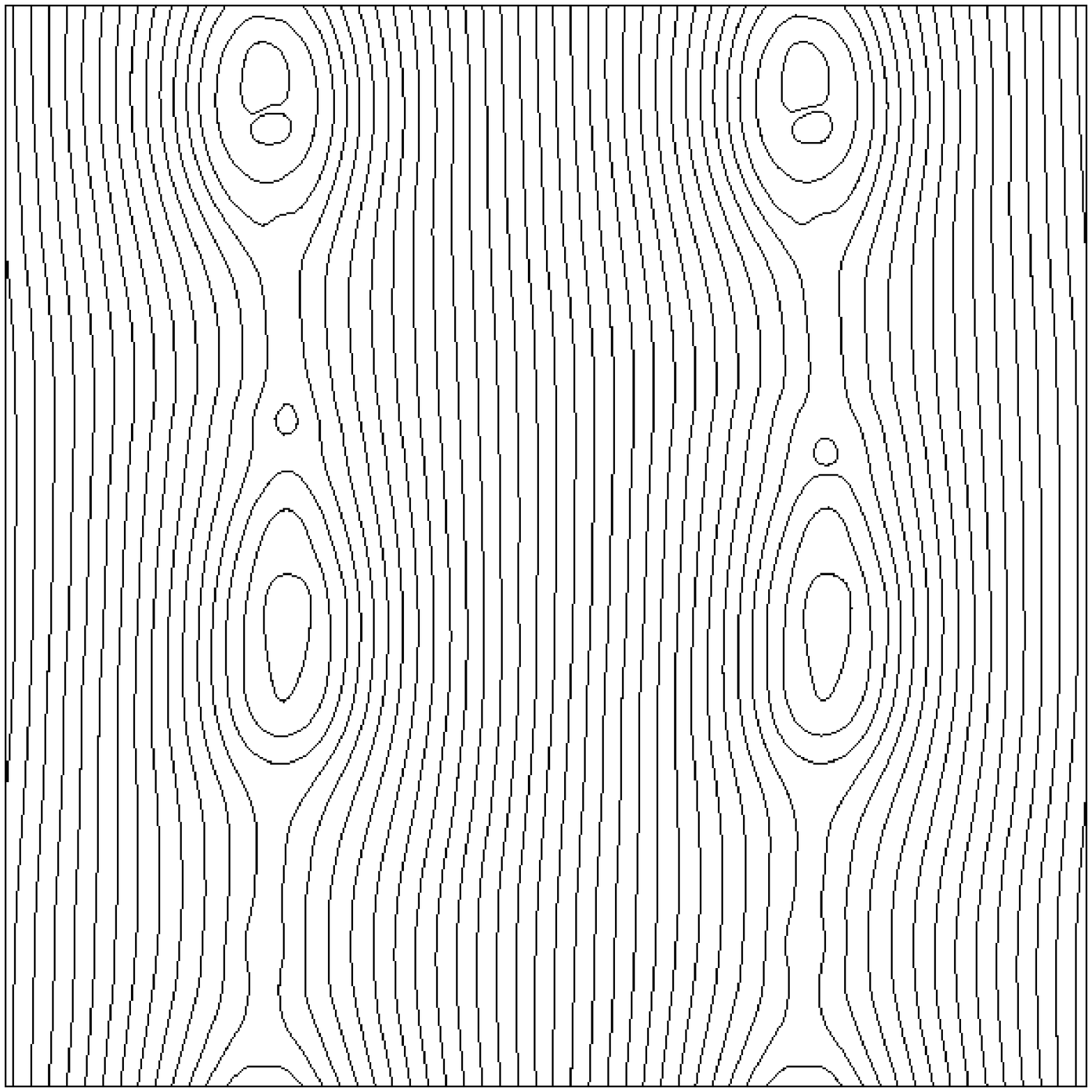} \hfill
\includegraphics*[width=1.5in]{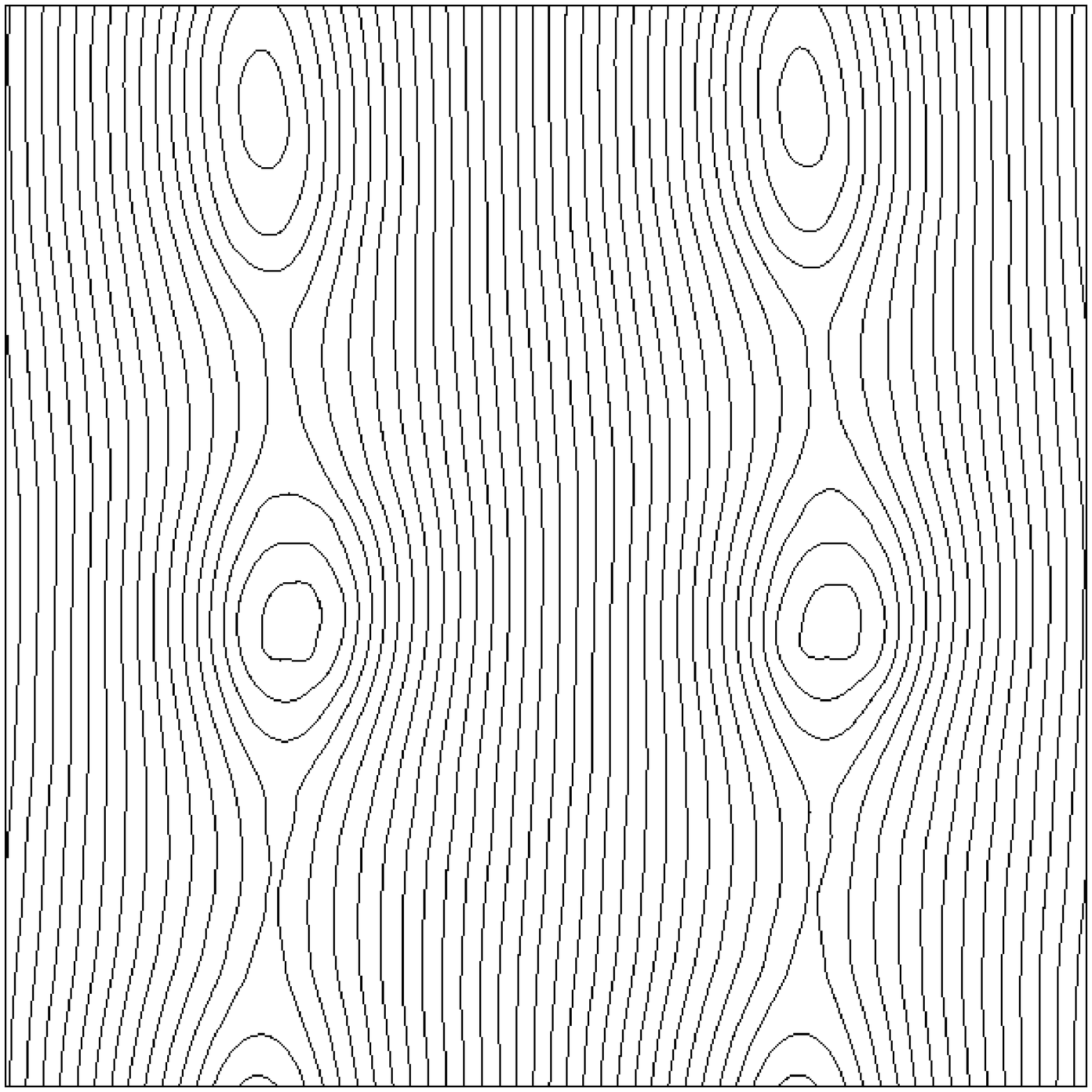} \\
\end{center}
\caption{Time evolution of the magnetic field lines using the CTU + CT 
integration algorithm.  Time increases from left to right and top to
bottom in normal reading order.  The contour levels of $A_z$ which are
plotted is uniform over the sequence of images at times $t=(0,~ 0.5,~
1.0,~ 1.5,~ 2.0,~ 2.5,~ 3.0,~ 3.5,~ 4.0)$.}
\label{fig:current_sheet}
\end{figure}

One aspect of this problem which is of particular interest is the
magnetic reconnection since it is a direct measure of the numerical
resistivity.  In figure {}\ref{fig:current_sheet} we present the time
evolution of the magnetic field lines.  From the magnetic field
geometry at time $t=0.5$ we see that, as one should expect, the
numerical resistivity is a function of the magnetic field orientation
with respect to the grid: magnetic fields dissipate, and reconnect
preferentially where the magnetic field orientation is oblique to the
grid.  Hence, we find the largest change in the magnetic field
structure at the nodal points of the transverse velocity. As
reconnection takes place, the magnetic energy is converted into
thermal energy (on time scales of the integration time step) which in
turn drives both compressional, and Alfv\'enic waves.  These waves
interact seeding more reconnection events.  By time $t=1$ a series of
magnetic islands have developed along the current sheets.  These
islands are free to move parallel to the local magnetic field
direction and by time $t=1.5$ some of the magnetic islands have
propagated toward the velocity anti-nodes and merged.  This process of
island formation, translation, and merging continues until there are
two magnetic field islands along each current sheet located
approximately at the velocity anti-nodes.

\par
This problem is also interesting in that it uses a very simple set of
initial conditions to test the ``robustness'' of the integration
algorithm.  The nonlinear dynamics which result from this problem lead
to strong compressions and rarefactions. It is important to maintain
the divergence free constraint as the topology of the field changes
during reconnection.  By either increasing $v_0$ or decreasing $P$
(and therefore $\beta$) the dynamics become increasingly difficult for
the integration algorithm to solve.  We  have found it a very useful test
to discriminate between algorithms.

%

\subsection{MHD Blast Wave}

As our final test problem we consider the explosion of a centrally
over pressurized region into a low pressure, low $\beta$ ambient
medium.  This problem has been studied by a number of authors
{}\cite{Zachary94,Balsara-Spicer,Londrillo-DelZanna00} and we've
chosen to use the parameters given by \cite{Londrillo-DelZanna00}.
The computational domain extends from $-0.5 \le x \le 0.5$ and $-0.5
\le y \le 0.5$.  The density $\rho=1$, the velocity ${\bf v=0}$, and the
magnetic field components $B_x = B_y = 10/\sqrt{2}$ and $B_z=0$.
Within a circle of radius $R = 0.125$ about the origin the gas
pressure $P=100$ and $\beta = 2 P/B^2 = 2$.  Outside of this circle,
the gas pressure $P=1$ and $\beta = 2\times 10^{-2}$.

\par
\begin{figure}[hbt]
\begin{center}
\includegraphics*[width=2.5 in]{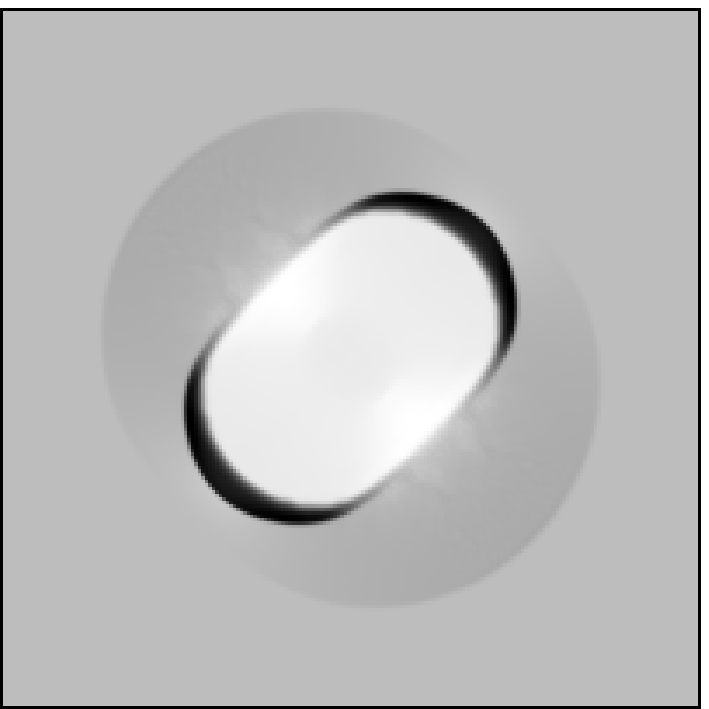} \hfill
\includegraphics*[width=2.5 in]{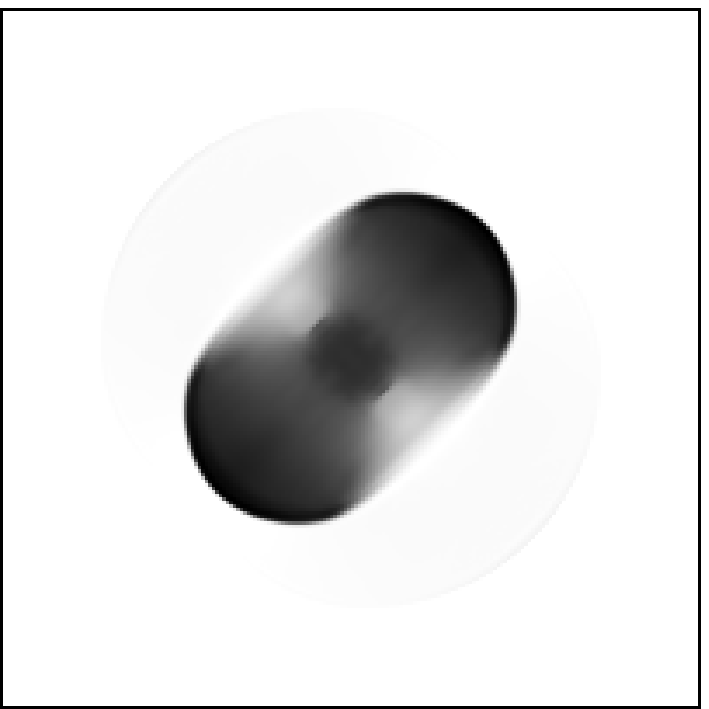} \\[0.25in]
\includegraphics*[width=2.5 in]{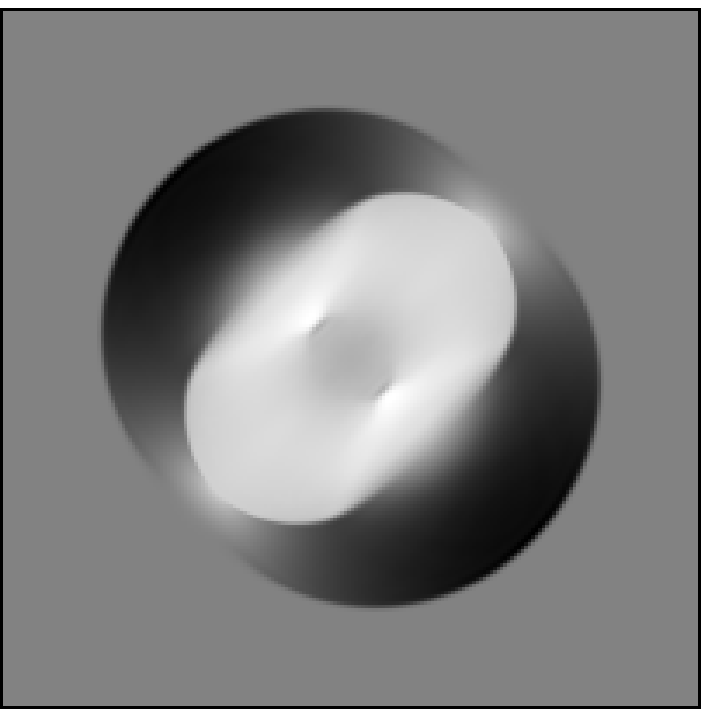} \hfill
\includegraphics*[width=2.5 in]{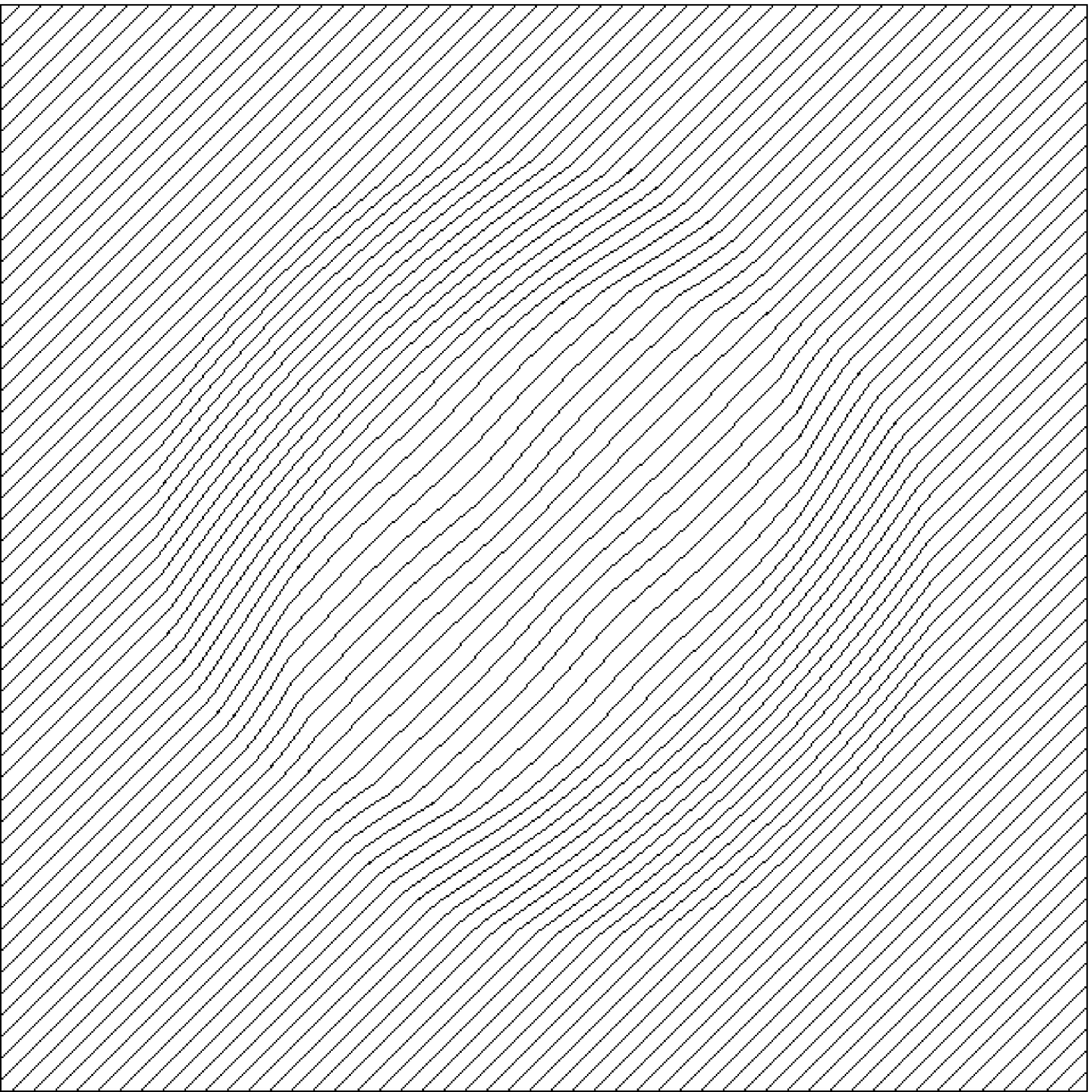} \\
\end{center}
\caption{Linearly scaled grey-scale images and the magnetic field lines 
of the evolved state (time=0.2) for the MHD blast wave problem.  The
density (\emph{top left}) ranges from 0.192 (white) - 3.31 (black).
The gas pressure (\emph{top right}) ranges from 1.0 (white) - 32.1
(black). The magnetic energy density (\emph{bottom left}) ranges from
23.5 (white) - 77.7 (black).}
\label{fig:mhd_blast}
\end{figure}

The solution to this problem at time $t=0.2$, using a $200 \times 200$
grid, is presented in figure \ref{fig:mhd_blast}.  The density image
shows two dense shells of gas which propagate parallel to the magnetic
field.  The outer surface of these shells is a slow-mode shock and the
inner surface is the contact surface separating the gas initially
inside and outside of the boundary surface. The maximum compression of
the gas is $3.3$ indicating that the slow mode shock is quite strong,
in agreement with what one might expect from the ratio of the gas
pressures, $P_{in}/P_{amb}=100$.  In the direction orthogonal to the
magnetic field, the magnetic pressure is the dominant player in the
dynamics, yet from the field lines we see that there is only a
moderate change in the geometry of the field.

\par
The solution presented in figure \ref{fig:mhd_blast} demonstrates that
the algorithm presented in this paper is both stable and accurate for
low-$\beta$ plasma problems involving strong MHD shock waves.  The
solutions also preserve the initial symmetry of the flow exceptionally
well despite the orientation of the magnetic field.  We find no
indication of grid related artifacts in the solution.

%

\section{Conclusion}
\label{sec:Conclusions}

In this paper we have demonstrated that the method of constrained
transport can be combined with finite volume integration algorithms in
a self consistent manner.  Consistency, however, implies that the
electric fields used in the CT update step and those used for evolving
the volume average magnetic fields are coupled.  This coupling has
direct consequences for the stability and accuracy of the integration
algorithm.  We have presented a general approach to constructing CT
algorithms, constructing and testing three.  Each of these CT
algorithms contained the novel property that for planar, grid-aligned
flows the solution would recover the 1D solution obtained with the
underlying integration algorithm.  These CT algorithms differed only
in their dissipation properties for truly multidimensional flows.
Through numerical experiments we have shown that the $\E^c_z$ CT
algorithm is well behaved leading to stable, non-oscillatory
solutions.  We have also noted how this algorithm can readily be
combined with other unsplit integration algorithms such as central
schemes, or wave propagation methods.

\par
We have shown that if the PPM integration algorithm is used for ideal
MHD, terms proportional to $\partial B_x/\partial x$ and $\partial
B_y/\partial y$, which in primitive variables are present only in the
induction equation, must be included in the calculation of the
``interface states''.  If these terms are neglected in the calculation
of the interface states, the integration algorithm is oscillatory for
the simple case of field loop advection.  We also presented two simple
\emph{gedanken} experiments to demonstrate why this result should be
expected.  A simple approach for including these source terms in the
calculation of the interface states is adopted and it has been
demonstrated to be accurate and stable for use with both the single
step and two step (CTU + CT) integration algorithm.

\par
Another result of this paper is the extension of the CTU integration
algorithm for ideal MHD based upon the CT integration algorithm.  We
showed that since the interface states are calculated using primitive
variables, the standard CTU procedure for updating the interface
states to the $1/2$ time step is missing terms which are proportional
to $\partial B_x/\partial x$ at $x$-interfaces and similarly for the
$y$-interface states.  These terms must be included as an additional
set of ``source terms'' so that the interface states are formally
advanced to the $1/2$ time step.  We also described how the CT
algorithms developed in this paper can be combined with the CTU
integration algorithm so as to maintain ${\bf\nabla\cdot B}=0$
throughout the integration time step.

\par
The CTU + CT integration algorithm presented in this paper for ideal
MHD has been thoroughly tested and some representative solutions have
been included here.  This algorithm combines the strong stability and
shock capturing characteristics of Godunov methods with the magnetic
flux conservation obtained via the CT method.  The integration
algorithm is conservative, uses a single step update algorithm, and is
second order accurate on smooth solutions.  These characteristics make
it ideally suited for use on either a statically or adaptively refined
mesh.  We note also that ``physical'' source terms such as an external
gravitational field, or Coriolis terms accounting for a rotating
reference frame can be readily incorporated into this integration
algorithm.  The resulting algorithm also retains the desirable
properties noted above, such as recovering the 1D solution for plane
parallel grid-aligned flows.  These details will be described in a
later paper.

\par
Lastly, while the algorithm presented in this paper has focused
solely on the two-dimensional case, the results presented here can be
extended to three dimensions.  This extension principally involves
modifications to the ``source terms'' involved in the calculation and
update of the interface states with transverse flux gradients.  The
actual details of this extension are beyond the scope of this paper and
will be presented elsewhere.

\section{Acknowledgements}

The authors would like to thank Charles Gammie, John Hawley, and Eve
Ostriker for discussion and comments on an early draft of this paper. We
thank Peter J. Teuben for contributions to the implementation of the
algorithms described here.  This work was supported by the National
Science Foundation ITR grant AST-0413788.  JS thanks the University of
Cambridge and the Royal Society for financial support during the course of
this work.

\appendix

\section{Linear Wave Right Eigenvectors}

In order to enable others to perform the linear wave convergence test
presented in section \ref{sec:linear_wave} and compare their results
in a quantitative manner, we include the numerical values for the
right eigenvectors here.  In the rotated coordinate system described
by equations (\ref{eq:coord_trans_x}) - (\ref{eq:coord_trans_z}) the
conserved variable vector
\beq
q = \left (
\begin{array}{c}
\rho \\
\rho v_1 \\
\rho v_2 \\
\rho v_3 \\
B_1 \\
B_2 \\
B_3 \\
E
\end{array}
\right ).
\eeq
The right eigenvectors (labeled according to their propagation
velocity) are given by
\beq
R_{\pm c_f} = \frac{1}{6\sqrt{5}} \left (
\begin{array}{c}
6 \\ 
\pm 12 \\ 
\mp 4\sqrt{2} \\ 
\mp 2 \\ 
 0 \\
8\sqrt{2} \\ 
4 \\ 
27 
\end{array}
\right ),~~~
%
%
R_{\pm c_a} = \frac{1}{3} \left (
\begin{array}{c}
 0 \\
 0 \\
\pm 1 \\ 
\mp 2 \sqrt{2} \\ 
 0 \\
-1 \\ 
2 \sqrt{2} \\ 
 0 \\
\end{array}
\right ).
\eeq
\beq
R_{\pm c_s} = \frac{1}{6\sqrt{5}} \left (
\begin{array}{c}
12 \\ 
\pm 6 \\ 
\pm 8\sqrt{2} \\ 
\pm 4 \\ 
 0 \\
-4\sqrt{2} \\ 
-2 \\ 
9 
\end{array}
\right ),~~~
%
%
R_{v_1} = \frac{1}{2} \left (
\begin{array}{c}
2 \\
2 \\
0 \\
0 \\
0 \\
0 \\
0 \\
1
\end{array}
\right ).
\eeq
%


\end{document}